# A New Cryptosystem Based On Hidden Order Groups

Amitabh Saxena and Ben Soh
Email: {asaxena, ben}@cs.latrobe.edu.au
Department of Computer Science and Computer Engineering
La Trobe University
VIC, Australia 3086

September 5, 2018

## Abstract

Let $G_1$ be a cyclic multiplicative group of order $n$. It is known that the Diffie-Hellman problem is random self-reducible in $G_1$ with respect to a fixed generator $g$ if $\phi(n)$ is known. That is, given $g, g^x \in G_1$ and having oracle access to a 'Diffie-Hellman Problem' solver with fixed generator $g$, it is possible to compute $g^{1/x} \in G_1$ in polynomial time (see theorem 3.2). On the other hand, it is not known if such a reduction exists when $\phi(n)$ is unknown (see conjuncture 3.1). We exploit this "gap" to construct a cryptosystem based on hidden order groups and present a practical implementation of a novel cryptographic primitive called an *Oracle Strong Associative One-Way Function* (O-SAOWF). O-SAOWFs have applications in multiparty protocols. We demonstrate this by presenting a key agreement protocol for dynamic ad-hoc groups.

## 1 Introduction

The problem of efficient key agreement in ad-hoc groups is a challenging problem, primarily because membership in such groups does not follow any specified pattern. We envisage an ad-hoc group as a broadcast group where members do not have one-to-one channels; rather they share the communication medium such that everyone within range is able to receive any broadcast message. An efficient group key agreement protocol in this scenario should satisfy the property that the shared group key is computable without interaction with the other members. Such protocols are often called *one-round* key agreement protocols where the only round consists of the initial key distribution phase. Two notable examples of one-round key agreement protocols are the classic two-party Diffie-Hellman key exchange [1] and the Joux tripartite key exchange using bilinear maps [2]. However, till date constructing a generalized one-round $n$-party key agreement protocol has remained a challenging and open problem. In this paper, we present the first practical example of a one-round key agreement protocol for arbitrary size groups. Although our construction enables the group key to be computed non-interactively, it comes with a caveat; a third party is required to do most of the computation.

We refer the reader to [3, 4] for a survey of key agreement protocols for ad-hoc groups. In the literature, most group key agreement protocols are classified in three categories (a) Centralized, (b) Distributed and (c) Fully Contributory. Our proposed method is fully contributory, yet it uses a central authority. We elaborate on this below.

The original two-party Diffie-Hellman key exchange [1] can be extended to fully contributory multi-party key exchange as demonstrated in [5] using the Group Diffie-Hellman (GDH) protocol. However, all protocols based on GDH require many rounds of sequential messages to be exchanged between members.

Centralized protocols, on the other hand have their own disadvantages; the central controller needs to maintain a large amount of state information for the groups it is managing. Our approach is to combine the two methods and design an efficient one-round key agreement protocol where the central controller does not maintain any state information.

Our protocol uses a central authority in computing the shared group key. However, the central authority is not responsible for key distribution and is only used as an "oracle" (i.e. a computing device) with public access. Users do not require secure channels in communicating with this oracle. Additionally, we provide a method to verify that the oracle is performing correctly. In our protocol, this oracle has some trapdoor information that can be efficiently used to compute partial public keys that are sent to users over an insecure public channel. Thus, our protocol can be directly converted into a de-centralized (or distributed) one simply by sharing this trapdoor information between a number of trusted authorities and allowing multiple "copies" of this oracle to function simultaneously. In effect, we present an entirely new model for secure group communication (see figure 1).

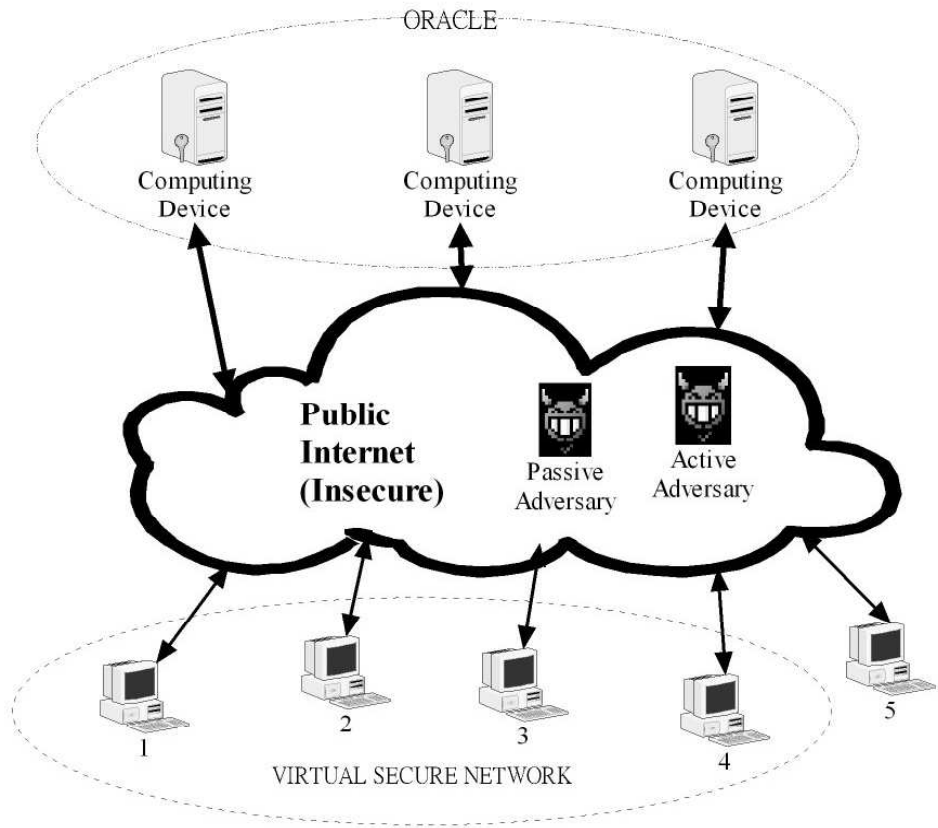

*In our model, secure group communication is facilitated by the Oracle. Assuming that public keys are known in advance, users can use this Oracle to compute a shared secret key independently of the other users such that no (active or passive) adversary has the ability to compute this key. Essentially the oracle is used as a "verifiable computing device" and the adversary as the communication medium.*

Figure 1: Secure group communication in our model.

Our basic idea arises due to the paper of Rabi and Sherman [6], where they described a cryptographic primitive called a *Strong Associative One-Way Function (SAOWF)*, and discussed as an application a one-round key agreement protocol in ad-hoc groups. In related work, Boneh and Silverberg also proposed a one-round key agreement protocol for ad-hoc groups based on a similar primitive called a *multilinear map* [7]. However, as of now no practical construction of either primitive is known. In this paper we extend the work of Rabi and Sherman and give a practical construction of a SAOWF under a restricted model of computation, namely *black-box computation*.

This paper is organized as follows. In section 2 we give some background and notation. We define

SAOWFs in section 2.1 and extend this definition to include black-box computation in section 2.4. Our construction is presented in section 4 and some applications are given in section 5. Finally, we discuss implementation issues in section 6.

## 2 Preliminaries

Around 1984, Rivest and Sherman suggested the idea of one-round key agreement in ad-hoc groups using a class of cryptographic primitives that they called *Associative One-Way Functions (AOWFs)* [8, 9]. Later in 1993, Rabi and Sherman suggested the use of AOWFs in digital signatures [10]. In subsequent work, Rabi and Sherman [6] gave an existence proof of complexity theoretic AOWFs under the $P \neq NP$ hypothesis. Other authors studied complexity theoretic AOWFs with respect to different properties such as low ambiguity, strong invertibility, totality and commutativity [11, 12, 13]. Finally, in [14], Hemaspaandra, Rothe and Saxena gave a complete characterization of complexity theoretic AOWFs.

In all the above works, however, the AOWFs considered are *complexity theoretic*, that is, they exhibit useful characteristics only in the *worst case* and not in the *average case*. Such constructions do not have much practical significance in the context of cryptography. In this work we focus on *cryptographic* AOWFs - that exhibit useful characteristics even in the average case. Additionally, we study only a small family of AOWFs, namely those that are commutative, total and strongly non-invertible. We call this the class of *Strong Associative One-Way Functions (SAOWFs)*.

### 2.1 Strong Associative One-Way Functions

Let $(\mathbb{G}, \star)$ be a finite abelian group. The mapping

$$\begin{aligned} f : \mathbb{G} \times \mathbb{G} &\mapsto \mathbb{G} \\ (A, B) &\mapsto A \star B \end{aligned}$$

has the following four properties (we use the notation $f(A, B)$ and $A \star B$ interchangeably):

P1. **Associativity**: $f(f(A, B), C) = f(A, f(B, C)) \ \forall A, B, C \in \mathbb{G}$.

P2. **Commutativity**: $f(A, B) = f(B, A) \ \forall A, B \in \mathbb{G}$.

P3. **Identity**: There exists a unique element $I \in \mathbb{G}$ such that $f(A, I) = A \ \forall A \in \mathbb{G}$. We say $I$ is the identity element. Denote by $\mathbb{G}^*$ the set $\mathbb{G} \backslash \{I\}$.

P4. **Inverses**: For each $A \in \mathbb{G}^*$, there exists a unique $B \in \mathbb{G}^*$ such that $f(A, B) = I$. We say $B$ is the inverse of $A$ and denote it by $A^{-1}$.

The above properties come for "free" in any abelian group. We now additionally want to enforce the following three properties on $(\mathbb{G}, \star)$:

P5. **Samplability**: Elements of $\mathbb{G}$ must be efficiently samplable.

P6. **Computability**: For all $A, B \in \mathbb{G}$, $f(A, B)$ must be efficiently computable.

P7. **Strong Non-Invertibility**: Let $A, B \xleftarrow{R} \mathbb{G}^*$ and $C \leftarrow f(A, B) \in \mathbb{G}$. Given $A, C$, computing $B = f(C, A^{-1})$ must be infeasible in the average case.

**Definition 2.1.** *We say that $f$ is a* Strong Associative One-Way Function *(SAOWF) if properties P1-P7 are satisfied.*[1]

---

[1] Most researchers differentiate between commutative and non-commutative SAOWFs [14]. For simplicity, we will enforce the commutativity property (P2) in our definition.

**Remark 2.2.** A SAOWF as defined above is analogous to a Group with Infeasible Inversion (GII) defined in [15].

Although SAOWFs have many applications as demonstrated in [6, 15, 16], exhibiting a practical construction of a SAOWF is still an open problem. We make a positive progress in this direction by presenting a practical black-box construction of a SAOWF.

We note that it is possible to construct a SAOWF $f$ under the $P \neq NP$ hypothesis if we replace "average case" by "worst case" in the statement of property P7 [13, 14]. However, for applications significant to cryptography we require property P7 to be defined in the average case. For completeness, we also define weak non-invertibility as follows.

P8. **Weak Non-Invertibility**: Let $C \xleftarrow{R} \mathbb{G}^*$. Given $C$, computing any pair $(A, B) \in \mathbb{G}^{*2}$ such that $C = f(A, B)$ must be infeasible in the average case.

**Definition 2.3.** *We say that $f$ is a* Weak Associative One-Way Function *(WAOWF) if properties P1-P6 and P8 are satisfied.*

The strong non-invertibility condition (P7) requires that for any $C \xleftarrow{R} image(f)$, inverting $f$ with respect to a **given** preimage $A$ must be infeasible in the average case. However, this condition does not say anything about weak non-invertibility (P8), which requires that computing **any** preimage of $C$ must be infeasible. In fact, the results of [17] prove that there exists an associative one-way function that is strongly non-invertible but not weakly non-invertible.[2]

Thus, a WAOWF may not be a SAOWF and vice-versa. In this work, we do not enforce the weak non-invertibility requirement. Rather, we allow the function to be weakly invertible. It turns out that our construction of a SAOWF is strongly non-invertible, yet it is weakly invertible.

Clearly, property P7 implies that computing inverses in $\mathbb{G}$ must be infeasible. Since the group $(\mathbb{G}, \star)$ is of finite order, the only way to achieve this is to keep the order of this group hidden. This is the main idea behind our construction.

## 2.2 Black-Box Constructions

Although the original objective of our research was to exhibit a practical construction of a SAOWF, in this work, we focus on a slightly different but related problem: exhibiting a practical *black-box* construction of a SAOWF by extending the definition of "computation" in property P6 to include *oracle computation*.

In our black-box model although the group $(\mathbb{G}, \star)$ is easily samplable, we we do not have access to the algorithm for computing $f$. Instead, access to the computing algorithm is only provided via a "black-box" with public access. This is illustrated in figure 2.

However, for a black-box construction to have any practical significance it must support (a) verifiable and (b) private computation as elaborated next.

## 2.3 PV-Oracles

In complexity theory, a black-box with public access is referred to as an *oracle*. In this work, we restrict ourselves to *constructible* oracles (i.e. oracles that can be constructed using some trapdoor), since we want our system to be practical. Additionally, to justify the use of a (constructible) oracle as one-way function in a cryptographic protocol, we must provide the same guarantees that a real function provides. Specifically, a real function is private and verifiable. We define similar properties for oracles. We will restrict ourselves to an oracle that computes a binary commutative function using two inputs.

---
[2] We note that the terminology used in this paper is slightly non-standard (but more intuitive). For instance, "weak non-invertibility" as defined here is simply referred to as "non-invertibility" in the literature [17]. Additionally, "weak" in the literature is used to refer to non-total functions [13]. However, since we are working in finite abelian groups, we can dispense off with definitions such as *honesty*, *non-commutativity* and *totality* used in [13, 14] for describing SAOWFs.

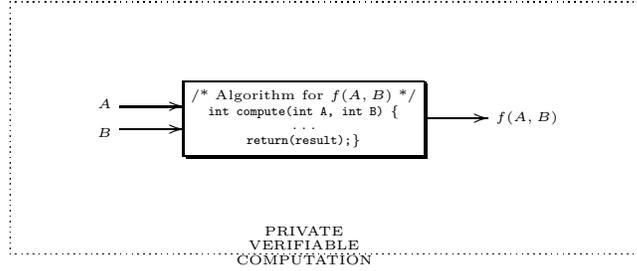

(a) A real computable function

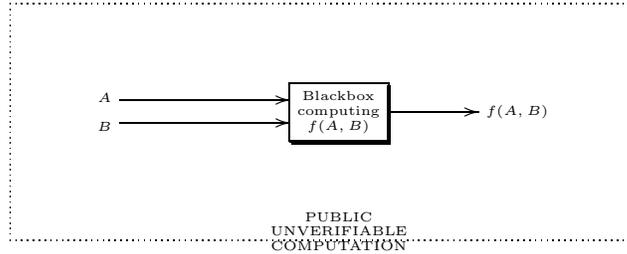

(b) A black-box with public access

Figure 2: Comparing a real and black-box computation.

**Verifiable Oracles.** Let $f$ be the binary commutative function computed by an oracle. We say that the oracle supports verifiable computation if for all $A, B \in domain(f)$ and all $C \in image(f)$, there exists a PPT verification algorithm Verify that outputs 1 if $C = f(A, B)$ and 0 otherwise. An oracle supporting verifiable computation is called a *Verifiable Oracle (V-Oracle)*. A V-Oracle is illustrated in figure 3.[3]

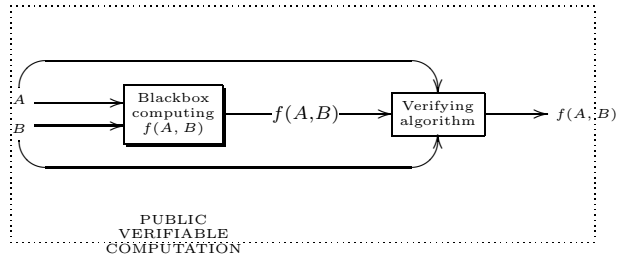

Figure 3: Public, Verifiable Black-box computation (V-Oracle).

**Private And Verifiable Oracles.** Let $f$ be the binary commutative function computed by a V-Oracle. We say that the V-Oracle supports private computation if the inputs and outputs of the computation can be *blinded* from the V-Oracle such that the blinding algorithm provides information theoretic [18] secrecy. Formally, there must exist two PPT algorithms Blind and Unblind as follows.

---

[3] As an example of a V-Oracle with one input, consider an existentially unforgeable signature scheme. The signing oracle is a V-Oracle since the signature can obviously be verified.

Blind is a randomized algorithm and takes as input an element $A \in domain(f)$ and outputs a tuple $(A', \sigma)$, where $A' \in domain(f)$ and the distributions $\{A\}$ and $\{A'\}$ are independent and identical. We say that Blind is the *Blinding Algorithm* and $\sigma$ is the *Unblinding Value*.

Unblind takes as input a tuple $(C', \sigma)$, where $C \in image(f)$. It outputs a value $C \in image(f)$ such that the following homomorphic property holds (We call Unblind the *Unblinding Algorithm*).

$$\Pr \left[ \begin{array}{rcl} A, B & \xleftarrow{R} & domain(f); \\ (A', \sigma) & \xleftarrow{R} & \mathsf{Blind}(A); \\ C & \leftarrow & \mathsf{Unblind}(f(A', B), \sigma) : \\ C & = & f(A, B) \end{array} \right] = 1 \qquad (1)$$

We call a V-Oracle supporting a private computation a *Private V-Oracle (PV-Oracle)*. See figure 4 for an illustration of a PV-Oracle.[4]

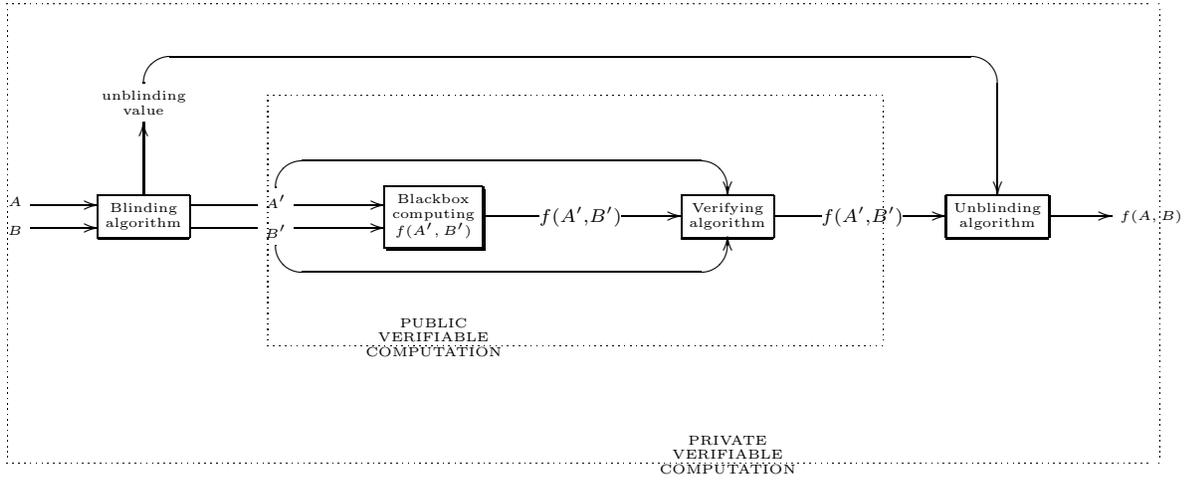

Figure 4: Private and Verifiable Black-box computation (PV-Oracle).

## 2.4 Oracle SAOWFs (O-SAOWFs)

We now extend the definition of computation in property P6 of section 2.1 to include computation by PV-Oracles. We call such a construction an *Oracle-SAOWF (O-SAOWF)* and formally define it below.

**Definition 2.4.** *A black-box construction of a SAOWF implemented using a PV-Oracle is called an Oracle-SAOWF (O-SAOWF). An O-SAOWF construction has four PPT "algorithms" as described below (we use quotes here because one of the algorithms PV-Compute is not a real algorithm in the usual sense; it involves a call to a PV-Oracle).*

**Setup** This is a randomized algorithm and takes in as input a security parameter $\tau$. It outputs the system parameters params for the group $(\mathbb{G}, \star)$ and a master key master-key.

**Sample** This is a randomized algorithm and takes in the parameter params. It outputs a uniformly selected element $A \xleftarrow{R} \mathbb{G}$ along with some auxiliary information $\sigma_A$, which we will call the *sampling information* in our construction.

---
[4]As an example of a PV-Oracle with one input, consider a RSA decryption oracle w.r.t. a given RSA public key. Information theoretic privacy for inputs to the decryption oracle can be achieved using Chaum's blinding technique [19].

**Compute** This algorithm takes as input the parameter params, the master key master-key and two values $A, B$. If $(A, B) \notin \mathbb{G}^2$, it sets $C \leftarrow I$ (recall that $I$ is the identity element). On the other hand, if $(A, B) \in \mathbb{G}^2$ it computes $C \leftarrow f(A, B) = A \star B$. The output is $C \in \mathbb{G}$. We assume that master-key acts like a trapdoor that enables computation of $f$.

Define a PV-Oracle $\mathcal{O}$ having access to master-key and implementing the compute algorithm. We assume that master-key is not known to anybody else. The fourth algorithm involves a call to this oracle.

**PV-Compute** This algorithm takes as input the parameter params and two elements $A, B \in \mathbb{G}$. It uses the Verify, Blind and Unblind algorithms defined in section 2.3 as sub-routines to compute $C \leftarrow f(A, B) = A \star B$ privately and verifiably by querying the PV-Oracle $\mathcal{O}$ that implements the Compute procedure. It outputs $C \in \mathbb{G}$.

### 2.4.1 Security Of O-SAOWFs

Assume that the PV-Compute algorithm performs correctly (that is, the Verify algorithm is correct and the Blind/Unblind algorithms provide information theoretic secrecy). Also assume that access to the Compute algorithm is available only in a black-box manner via oracle $\mathcal{O}$ that knows the parameter master-key. We can then define the security of the O-SAOWF as follows. We say that a PPT algorithm $\mathcal{A}$ breaks the O-SAOWF if it is able to "strongly invert" the O-SAOWF and compute inverses in $\mathbb{G}$ having only black-box access to the Compute algorithm. We call this the *Group Inversion Problem (GIP$_\mathbb{G}$)*. Formally, the advantage of $\mathcal{A}$ in solving GIP$_\mathbb{G}$ is defined as

$$\text{GIP-Adv}_\mathcal{A}(\tau) = \Pr \left[ \begin{array}{l} \mathcal{A}^{\mathcal{O}(\text{Compute}(),\ \text{master-key})}(P, \text{params}) \to P^{-1} : \\ \\ (\text{params}, \text{master-key}) \xleftarrow{R} \text{Setup}(\tau), \\ (P, \sigma_P) \xleftarrow{R} \text{Sample}(\text{params}) \end{array} \right] \quad (2)$$

**Definition 2.5.** *We say that algorithm $\mathcal{A}$ $(k_\mathcal{O}, \delta, \epsilon)$-breaks the O-SAOWF $f$ if $\mathcal{A}$ runs at most time $\delta$; $\mathcal{A}$ makes at most $k_\mathcal{O}$ adaptive queries to the oracle $\mathcal{O}$ implementing the Compute algorithm; and GIP-Adv$_\mathcal{A}(\tau)$ is at least $\epsilon$. Alternatively we say that the O-SAOWF is $(k_\mathcal{O}, \delta, \epsilon)$-secure under an adaptive attack if no such algorithm $\mathcal{A}$ exits.*

It is clear that a black-box SAOWF $f$ where we extend the definition of computation in property P6 to include computation by PV-Oracles, is identical to a "real" computable SAOWF $f$ in terms of functionality. However, until now it had been an open question to present even a black-box construction of SAOWFs using PV-Oracles. In this work, we present the first practical construction of a black-box SAOWF based on a PV-Oracle. In other words, our construction allows private (in the information-theoretic sense) and verifiable computation.[5]

**Remark 2.6.** It should be noted that the above model of an O-SAOWF $f$ that allows black-box computation of the group operation $\star$ on $\mathbb{G}$ using a PV-Oracle is different from a *black-box group*, a notion introduced by Babai and Szemerédi [20] (see also [21]), where access to the entire group $(\mathbb{G}, \star)$ is provided through black-box routines and the representation of group elements is opaque. In contrast, the above model is an example of a *semi black-box group*, since the representation of group elements is not opaque and certain operations like blinding/unblinding, sampling and verification of composition can be done outside of the black-box.

## 3 The Underlying Primitives

In this section, we give a brief overview of the two main underlying primitives of our construction: (i) composite order bilinear maps, and (ii) the Paillier cryptosystem.

---
[5]It is noteworthy that our construction of a black-box SAOWF using a PV-Oracle also serves an existence proof of real computable SAOWFs in a way because we achieve almost identical functionality using a black-box construction.

## 3.1 Bilinear Maps

Let $G_1$ and $G_2$ be two cyclic multiplicative groups both of the same order $n$ such that computing discrete logarithms in $G_1$ and $G_2$ is intractable. A bilinear pairing is a map $\hat{e}: G_1 \times G_1 \mapsto G_2$ that satisfies the following properties:

1. *Bilinearity*: $\hat{e}(a^x, b^y) = \hat{e}(a,b)^{xy}$ $\forall a,b \in G_1$ and $x,y \in \mathbb{Z}_n$.

2. *Non-degeneracy*: If $g$ is a generator of $G_1$ then $\hat{e}(g,g)$ is a generator of $G_2$.

3. *Computability*: The map $\hat{e}$ is efficiently computable.

The above properties also imply:

$$\hat{e}(ab, c) = \hat{e}(a,c) \cdot \hat{e}(b,c) \; \forall a,b,c \in G_1$$

$$\hat{e}(a, bc) = \hat{e}(a,b) \cdot \hat{e}(a,c) \; \forall a,b,c \in G_1$$

Additionally, we assume that it is easy to *sample* elements from $G_1$. In a practical implementation, the group $G_1$ is the set of points on an elliptic curve and $G_2$ is the multiplicative subgroup of a finite field. The map $\hat{e}$ is derived either from the modified Weil pairing [22, 23] or the Tate pairing [24]. We will assume that the smallest prime factor of $n$ is $\geq 2^{171}$ so that the fastest algorithm for computing discrete logarithms in $G_1$ (Pollard's rho method [25, p.128]) takes $\geq 2^{85}$ iterations [22].

### 3.1.1 Problems in Bilinear Maps

It is clear that irrespective of whether $n$ is prime or composite, both $G_1$ and $G_2$ have generators. Fix some generator $g$ of $G_1$ and define the following problems.

**Computational Diffie-Hellman Problem [CDHP$_{(g,G_1)}$]:** Given $g^x, g^y \in G_1$, output $g^{xy} \in G_1$.

**Decision Diffie-Hellman Problem [DDHP$_{(g,G_1)}$]:** Given $g^x, g^y, g^z \in G_1$, output 1 if $z = xy \in \mathbb{Z}_n$; otherwise output 0.

**Inverse Diffie-Hellman Problem [IDHP$_{(g,G_1)}$]:** Given $g^x \in G_1$ for some $x \in \mathbb{Z}_n^*$, output $g^{1/x} \in G_1$.

The following result was noted by Joux and Nguyen [26].

**Lemma 3.1.** *DDHP$_{(g,G_1)}$ (the decision Diffie-Hellman problem) is easy.*

*Proof.* Clearly, from the properties of the mapping, $z = xy \in \mathbb{Z}_n$ if and only if $\hat{e}(g, g^z) = \hat{e}(g^x, g^y)$. Thus, solving DDHP$_{(g,G_1)}$ is equivalent to computing the mapping $\hat{e}$ twice. □

The next theorem shows that the computational Diffie-Hellman problem is random self-reducible in the group $G$ if $\phi(n)$ is known.

**Theorem 3.2.** *IDHP$_{(g,G_1)}$ $\Rightarrow$ CDHP$_{(g,G_1)}$ if $\phi(n)$ is known.*

*Proof.* We must show that given an IDHP$_{(g,G_1)}$ instance $g^x \in G_1$ for some $x \in \mathbb{Z}_n^*$ and access to a CDHP$_{(g,G_1)}$ oracle, we can efficiently compute $g^{1/x} \in G_1$. This follows from the following facts.

1. **Fact.** From Euler's theorem [25, p.69] we know that $\forall u \in \mathbb{Z}_n^*$ $u^{\phi(n)} \equiv 1 \mod n$. Equivalently, $u^{\phi(n)-1} \equiv 1/u \mod n$.

2. **Fact.** Given any pair $g^u, g^v \in G_1$ for arbitrary $u, v \in \mathbb{N}$ we can use the CDHP$_{(g,G_1)}$ oracle to compute $g^{uv} \in G_1$

3. **Fact.** Given any value $g^u \in G_1$, we can use the CDHP$_{(g,G_1)}$ oracle to compute $g^{u^{2^i}}$ for any $i \in \mathbb{N}$ by the "repeated squaring" method (see [27, p.23] for an example).

Therefore, from $g^x$ we can efficiently compute $h = g^{x^{\phi(n)-1}} \in G_1$ using the CDHP$_{(g,G_1)}$ oracle and the "repeated squaring and multiply" algorithm of [25, p.71] (via facts 2 and 3). Then from fact 1, $h = g^{1/x}$, and thus, $h$ is the required solution. □

Although theorem 3.2 says that IDHP$_{(g,G_1)} \Rightarrow$ CDHP$_{(g,G_1)}$ if $\phi(n)$ is known, it is not clear if the same reduction holds when $\phi(n)$ is unknown. In light of this, we make the following hypothesis, necessary for the security of our construction.

**Conjuncture 3.1.** *IDHP$_{(g,G_1)} \not\Rightarrow$ CDHP$_{(g,G_1)}$ if $\phi(n)$ is unknown.*

### 3.1.2 BDH Parameter Generator

We will further assume that $n = |G_1| = |G_2| = pq$ where $p, q$ are large primes such that given the product $n = pq$, factoring $n$ is intractable. We refer the reader to [28] for details on generating composite order bilinear maps for any given $n$ that is square free.

Using the idea of [23], we define a Bilinear Diffie-Hellman (BDH) parameter generator as a randomized PPT algorithm $\mathcal{BDH}$ that takes a single parameter $\tau \in \mathbb{N}$ and outputs a tuple $(\hat{e}, G_1, G_2, p, q)$ such that $p, q$ are distinct primes of $\tau$ bits each, $G_1$, $G_2$ are two cyclic multiplicative groups of the same order $pq$, and $\hat{e}: G_1 \times G_1 \mapsto G_2$ is a bilinear mapping as defined in section 3.1.

For any PPT algorithm $\mathcal{A}$, denote by CDHP-Adv$_\mathcal{A}(\tau)$, the advantage of $\mathcal{A}$ in solving CDHP$_{(g,G_1)}$ for some security parameter $\tau$. Formally,

$$\text{CDHP-Adv}_\mathcal{A}(\tau) = \Pr \left[ \begin{array}{l} \mathcal{A}(\hat{e}, n, G_1, G_2, g, u, v) = g^{xy} : \\ \\ (\hat{e}, G_1, G_2, p, q) \xleftarrow{R} \mathcal{BDH}(\tau) \text{ s.t. } |G_1| = |G_2| = pq, \\ n = pq, \quad g \xleftarrow{R} G_1 \text{ s.t. } \langle g \rangle = G_1, \quad (x, y) \xleftarrow{R} \mathbb{Z}_n^2, \quad u = g^x, \quad v = g^y \end{array} \right] \quad (3)$$

Similarly, we denote by IDHP-Adv$_\mathcal{A}(\tau)$ the advantage of $\mathcal{A}$ in solving IDHP$_{(g,G_1)}$. Formally,

$$\text{IDHP-Adv}_\mathcal{A}(\tau) = \Pr \left[ \begin{array}{l} \mathcal{A}(\hat{e}, n, G_1, G_2, g, u) = g^{1/x} : \\ \\ (\hat{e}, G_1, G_2, p, q) \xleftarrow{R} \mathcal{BDH}(\tau) \text{ s.t. } |G_1| = |G_2| = pq, \\ n = pq, \quad g \xleftarrow{R} G_1 \text{ s.t. } \langle g \rangle = G_1, \quad x \xleftarrow{R} \mathbb{Z}_n^*, \quad u = g^x \end{array} \right] \quad (4)$$

We will make the following two assumptions for all our constructions.

**Diffie-Hellman Assumption:** The computation Diffie-Hellman problem (CDHP$_{(g,G_1)}$) is intractable. In other words, for all PPT algorithms $\mathcal{A}$, CDHP-Adv$_\mathcal{A}(\tau)$ is a negligible function of $\tau$.

**Inverse Diffie-Hellman Assumption:** The inverse Diffie-Hellman problem (IDHP$_{(g,G_1)}$) is intractable. In other words, for all PPT algorithms $\mathcal{A}$, IDHP-Adv$_\mathcal{A}(\tau)$ is a negligible function of $\tau$.

## 3.2 The Paillier Cryptosystem

Our idea of constructing the O-SAOWF is to use an oracle as a "Diffie-Hellman problem" solver in the bilinear group $G_1$ of composite order $n$. Since the only known way to solve the Diffie-Hellman problem is to compute discrete logarithms, we provide the discrete logarithms to the oracle in an encrypted form using an asymmetric cryptosystem. The requirement here is that the encryption algorithm **E** must possess the following multiplicative homomorphic property: for any messages $m_1, m_2 \in \mathbb{Z}_n^*$, given $\{\mathbf{E}(m_1), m_2\}$ or $\{m_1, \mathbf{E}(m_2)\}$, it must be possible to compute $\mathbf{E}(m_1 m_2 \bmod n)$ directly without knowing the corresponding decryption algorithm **D**. The Paillier cryptosystem [29] has this property.[6]

---
[6]Although this property is necessary, it is not sufficient; the RSA [30] and Rabin [31] cryptosystems also have this property. However, our construction based on RSA or Rabin is insecure.

The following facts form the basis of the Paillier cryptosystem. Let $n = pq$, where $p, q$ are distinct odd primes. Let $\lambda = \text{lcm}(p-1, q-1)$ and $\phi(n) = (p-1)(q-1)$.

1. **Fact.** $|\mathbb{Z}_{n^2}^*| = n\phi(n)$

2. **Fact.** For all $w \in \mathbb{Z}_{n^2}^*$ it is true that $w^{n\lambda} \equiv 1 \bmod n^2$ and $w^\lambda \equiv 1 \bmod n$.

3. **Fact.** For all $w \in \mathbb{Z}_{n^2}^*$ it is true that $(w^\lambda \bmod n^2) \equiv 1 \bmod n$. Thus the mapping $L : \mathbb{Z}_{n^2}^* \mapsto \mathbb{Z}_n$, where $L(w) = \frac{(w^\lambda \bmod n^2) - 1}{n}$ is well defined.

We are now ready to describe the Paillier cryptosystem (see [29] for details).

**Key Generation:** Generate $p, q \xleftarrow{R} \mathbb{N}$, where $p, q$ are large distinct primes. Set $n \leftarrow pq$ and $\lambda \leftarrow \text{lcm}(p-1, q-1)$. Generate $t \xleftarrow{R} \mathbb{Z}_{n^2}^*$ such that the order of $t$ is a non-zero multiple of $n$. This can be done by checking that $L(t^\lambda \bmod n^2)$ is invertible in $\mathbb{Z}_n$. The public key is $(t, n)$ and the private key is $(\lambda, n)$.

For convenience in this paper, we will use the notation $\mathbf{E}, \mathbf{D}$ to denote the encryption and decryption functions respectively for some fixed parameters $(\lambda, t, n)$ whenever the parameters are clear from the context.

**Encrypt:** To encrypt a message $m \in \mathbb{Z}_n$, generate random $r \xleftarrow{R} \mathbb{Z}_n^*$ and set

$$c \leftarrow \mathbf{E}(m) = t^m r^n \bmod n^2$$

The ciphertext is $c \in \mathbb{Z}_{n^2}^*$.

**Decrypt:** To decrypt, compute

$$m \leftarrow \mathbf{D}(c) = \frac{L(c^\lambda \bmod n^2)}{L(t^\lambda \bmod n^2)} \in \mathbb{Z}_n$$

### 3.2.1 Homomorphic Properties

The Paillier cryptosystem has the following homomorphic properties [29].

1. Plaintext multiplication:

$$\forall m_1, m_2 \in \mathbb{Z}_n \quad \mathbf{D}(\mathbf{E}(m_1)^{m_2} \bmod n^2) = \mathbf{D}(\mathbf{E}(m_2)^{m_1} \bmod n^2) = m_1 m_2 \bmod n$$

2. Self Blinding:

$$\forall m \in \mathbb{Z}_n \ \forall r \in \mathbb{N} \quad \mathbf{D}(\mathbf{E}(m) r^n \bmod n^2) = m$$

The semantic security of the above encryption scheme is proved under the *Decision Composite Residuosity Assumption* (DCRA) [29], which states that the following problem is hard unless the factors of $n$ are known.

**Decision Composite Residousity Problem [DCRP$_n$]** Given $x \xleftarrow{R} \mathbb{Z}_{n^2}^*$, output 1 if $\exists y \in \mathbb{Z}_{n^2}^*$ s.t. $x \equiv y^n \pmod{n^2}$ otherwise output 0.

The DCRA is a stronger assumption than factoring [29]. See [32, 33] for a discussion on the bit-security of the Paillier cryptosystem.

## 4 Our O-SAOWF Construction

Our construction will describe the four algorithms Setup, Sample, Compute and PV-Compute defined in section 2.4. For clarity of presentation, we give the construction in stages. First we describe the underlying primitives of our construction and any necessary security assumptions. Next, we describe the Setup procedure and elaborate on the structure of the group $(\mathbb{G}, \star)$ defined by params before describing the remaining algorithms.

## 4.1 Setup

This algorithm generates the system parameters. The input is a single parameter $\tau \in \mathbb{N}$.

1. Use the BDH parameter generator $\mathcal{BDH}$ of section 3.1.2 to output $(\hat{e}, G_1, G_2, p, q) \leftarrow \mathcal{BDH}(\tau, 2)$, where $p, q$ are large distinct primes of $\approx \tau$ bits each, $G_1, G_2$ are descriptions of two groups both of order $pq$ and $\hat{e} : G_1 \times G_1 \mapsto G_2$ is a bilinear map (of section 3.1). Then pick a generator $g \xleftarrow{R} G_1$.

2. Set $n \leftarrow pq$ and $\lambda \leftarrow \text{lcm}(p-1, q-1)$. Then generate an element $t \xleftarrow{R} \mathbb{Z}_{n^2}^*$ such that the order of $t$ is a non-zero multiple of $n$. The pair $(t, n)$ is the public key for the Paillier cryptosystem. The corresponding private key is $(\lambda, n)$. We will denote the corresponding encryption and decryption algorithms by $\mathbf{E}$ and $\mathbf{D}$ respectively.

3. Generate $\alpha, r \xleftarrow{R} \mathbb{Z}_n^*$. Then set $h \leftarrow g^\alpha \in G_1$ and $\beta \leftarrow \mathbf{E}(\alpha) = t^\alpha r^n \in \mathbb{Z}_{n^2}^*$.

4. Output params $\leftarrow (\hat{e}, G_1, G_2, g, t, n, h, \beta)$ and master-key $\leftarrow \lambda$.

Recall that the Compute algorithm requires as input the parameter master-key and is accessible only as a black-box routine via oracle $\mathcal{O}$ that implements this algorithm. The value master-key is sent to $\mathcal{O}$ via a secure channel and the value params is made public.

## 4.2 Description Of $(\mathbb{G}, \star)$

From params, the tuple $(\hat{e}, G_1, G_2, g, t, n)$ defines the structure of the group $(\mathbb{G}, \star)$ and the pair $(h, \beta)$ represents a random element of this group. We now describe the structure of this group.

1. Consider the set $\mathbb{S} \subsetneq G_1$ defined as

$$\mathbb{S} = \{x | x = g^y \text{ for some } y \in \mathbb{Z}_n^*\}$$

Clearly, $|\mathbb{S}| = \phi(n) = |\mathbb{Z}_n^*|$ and $\mathbb{S}$ is exactly the set of elements of $G_1$ having order $n$.

2. Define the set $\mathbb{G} \subsetneq \mathbb{S} \times \mathbb{Z}_{n^2}^*$ as
$$\mathbb{G} = \{(x, y) | x = g^{\mathbf{D}(y)}\} \quad (5)$$

and define a binary operation $\star$ on $\mathbb{G}$ using the multi-valued mapping

$$\begin{aligned} f : \mathbb{G} \times \mathbb{G} &\mapsto \mathbb{G} \\ (A, B) &\mapsto A \star B \end{aligned}$$

as follows. Let $A = (x_A, y_A)$ and $B = (x_B, y_B)$. Then $A \star B = (x_C, y_C)$, where

$$x_C \leftarrow x_A^{\mathbf{D}(y_B)} = g^{\mathbf{D}(y_A)\mathbf{D}(y_B)} = x_B^{\mathbf{D}(y_A)} \in G_1 \quad (6)$$

$$y_C \leftarrow \mathbf{E}(\mathbf{D}(y_A)\mathbf{D}(y_B) \bmod n) \in \mathbb{Z}_{n^2}^* \quad (7)$$

Thus, $x_C = g^{\mathbf{D}(y_C)}$ and therefore $(x_C, y_C) \in \mathbb{G}$.

3. Finally, define an equivalence relation $\sim$ on $\mathbb{G}$ as follows. For any $A, B \in \mathbb{G}$, where $A = (x_A, y_A)$ and $B = (x_B, y_B)$, we say that $A \sim B$ if and only if $x_A = x_B$. This relation is symmetric, reflexive and transitive. Thus, it indeed forms an equivalence relation.

We state without proof the following lemmas (which can be easily verified):

**Lemma 4.1.** *For any $A, B \in \mathbb{G}$, it is true that $A \star B \sim B \star A$. That is, the relation $\sim$ transforms $\star$ into an commutative operation over $\mathbb{G}$.*

**Lemma 4.2.** *For any $A, B, C \in \mathbb{G}$, it is true that $(A \star B) \star C \sim A \star (B \star C)$. That is, the relation $\sim$ transforms $\star$ into an associative operation over $\mathbb{G}$.*

For any $A \in \mathbb{G}$, denote by $[A] \subsetneq \mathbb{G}$ the *equivalence class* of $A$ with respect to the relation $\sim$. Therefore we can define an equivalence class $[I] \subsetneq \mathbb{G}$ as follows:

$$[I] = \{X | X \sim (g, t) \sim (g, \mathbf{E}(1))\}$$

**Lemma 4.3.** *For any $[A] \subsetneq \mathbb{G}$, there exists a unique $[B] \subsetneq \mathbb{G}$ such that $[A] \star [B] = [I]$. Additionally, $[A] \star [I] = [A]$.*

It is clear from the above lemmas that the relation $\sim$ transforms the equivalence classes of $\mathbb{G}$ into an abelian group with respect to the binary operation $\star$. The order of this group ($\phi(n)$) is effectively hidden from anyone who does not know the factors of $n$.

For any $[A] \subsetneq \mathbb{G}$, let the symbol $[A]^i$ denote $[A] \star [A] \star \ldots [A]$ ($i$ times). The inverse of $[A]$ is denoted by $[A]^{-1}$. It can be trivially verified that the following are also true.

$$\left. \begin{array}{rcl} [A]^i \star [A]^j & = & [A]^{i+j} \\ ([A]^i)^j & = & [A]^{ij} \\ [A] \star [A]^{-1} & = & [A]^0 = [I] \\ ([A]^i \star [B]^j)^k & = & [A]^{ik} \star [B]^{jk} \end{array} \right\} \begin{array}{l} \forall \, [A], [B] \subsetneq \mathbb{G} \\ \forall \, i, j, k \in \mathbb{Z} \end{array}$$

We will slightly abuse notation and denote the equivalence class $[A]$ by $A$. We will use $=$ instead of $\sim$ to indicate that we are working with equivalence classes. For any $j$ given elements $A_1, A_2, \ldots A_j \in \mathbb{G}$, we denote $A_1 \star A_2 \star \ldots A_j$ by

$$\prod_{i=1}^{j} A_i$$

## 4.3 Properties Of $(\mathbb{G}, \star)$

We now enumerate some important properties of the group $(\mathbb{G}, \star)$.

1. **Samplability:** $\mathbb{G}$ is efficiently samplable. To sample from $\mathbb{G}$, first generate random $\sigma \stackrel{R}{\leftarrow} \mathbb{Z}_n^*$. Then set $x \leftarrow g^\sigma \in G_1$ and $y \leftarrow \mathbf{E}(\sigma) \in \mathbb{Z}_{n^2}^*$. We see that $(x, y) \in \mathbb{G}$. In this case we call $\sigma$, the *sampling information* of $(x, y)$. When we say that $A \in \mathbb{G}$ has been sampled by us, we imply that the sampling information of $A$ is known. The sampling information acts like a trapdoor in our construction.

2. **Trapdoor Computability:** Let $A, B \in \mathbb{G}$ be given. Anyone who has sampled either one of $A$ or $B$ can compute $A \star B$ efficiently as follows:

   Let $A = (x_A, y_A)$ and $B = (x_B, y_B)$ be given. Additionally, we are given $\sigma_A \in \mathbb{Z}_n^*$, the sampling information of $A$. That is, $x_A = g^{\sigma_A} \in G_1$ and $y_A = \mathbf{E}(\sigma_A) \in \mathbb{Z}_{n^2}^*$. To compute $A \star B$, first generate random $r \stackrel{R}{\leftarrow} \mathbb{Z}_n^*$. Then set $x \leftarrow x_B^{\sigma_A} \in G_1$ and $y \leftarrow y_B^{\sigma_A} \cdot r^n \in \mathbb{Z}_{n^2}^*$.

   Therefore, $x = x_B^{\mathbf{D}(y_A)}$ and due to the homomorphic properties of the Paillier cryptosystem, we find that $y = \mathbf{E}(\sigma_A \mathbf{D}(y_B) \bmod n) = \mathbf{E}(\mathbf{D}(y_A)\mathbf{D}(y_B) \bmod n)$. Thus, $(x, y) = A \star B$.

3. **Trapdoor Strong Invertibility and Exponentiation:** Let $A, B \in \mathbb{G}$ be given. Anyone who has sampled $A \in \mathbb{G}$ can also compute $A^{-1} \star B$ because if $\sigma_A \in \mathbb{Z}_n^*$ is the sampling information for $A$ then $\sigma_A^{-1} \in \mathbb{Z}_n^*$ is the sampling information for $A^{-1}$. Also, for any $i \in \mathbb{Z}$, the sampling information for $A^i \in \mathbb{G}$ is $(\sigma_A)^i \in \mathbb{Z}_n^*$.

4. **Non-computability:** Let $A, B \in \mathbb{G}$ be given. Anyone who has *not sampled* at least one of $\{A, B, A^{-1}, B^{-1}\}$ cannot compute $A \star B$ without knowledge of $\lambda$.

5. **Strong Non-invertibility:** Let $A, B \in \mathbb{G}$ be given. Anyone who has *not sampled* at least one of $\{A, A^{-1}\}$ cannot compute $A^{-1} \star B$ without knowledge of $\lambda$.

6. **Indistinguishability:** Let $(x,y) \in G_1 \times \mathbb{Z}_{n^2}^*$ be given. It is infeasible to decide if $(x,y) \stackrel{?}{\in} \mathbb{G}$ without knowledge of $\lambda$.

7. **Black-Box Computability:** Let $A, B \in \mathbb{G}$ be given. Anyone knowing $\lambda$ has the ability to compute $A \star B$ using equations 7 and 6.

8. **Black-Box Distinguishability:** Let $(x,y) \in G_1 \times \mathbb{Z}_{n^2}^*$ be given. Anyone knowing $\lambda$, also has the ability to decide if $(x,y) \stackrel{?}{\in} \mathbb{G}$ by virtue of equation 5.

## 4.4 A Concrete O-SAOWF Construction

We now describe a concrete construction of an O-SAOWF under definition 2.4. In addition to the four main algorithms Setup, Sample, Compute, PV-Compute and the three algorithms Verify, Blind and Unblind used as subroutines in PV-Compute, our construction has four 'auxiliary' algorithms Verify-In-Group, Verify-Not-In-Group, TD-Exponentiate and V-Compute. Thus, our construction has a total of eleven algorithms. The Setup algorithm is described in section 4.1 while the Sample algorithm is described in section 4.3, item 1.

A-1.
> **Setup**
> *Input:* $\tau \in \mathbb{N}$
> *Step-1.* Generate $\{\hat{e}, G_1, G_2, g, t, n, h, \beta, \lambda\}$ as described in section 4.1.
> *Step-2.* Set params $\leftarrow (\hat{e}, G_1, G_2, g, t, n, h, \beta)$ and master-key $\leftarrow \lambda$.
> *Output:* (params, master-key)

A-2.
> **Sample**
> *Input:* params
> *Step-1.* Generate $\sigma_A, r \stackrel{R}{\leftarrow} \mathbb{Z}_n^*$
> *Step-2.* Set $x_A \leftarrow g^{\sigma_A} \in G_1$; $y_A \leftarrow t^{\sigma_A} r^n \bmod n^2 = \mathbf{E}(\sigma_A) \in \mathbb{Z}_{n^2}^*$
> *Step-3.* Set $A \leftarrow (x_A, y_A) \in \mathbb{G}$
> *Output:* $(A, \sigma_A) \in \mathbb{G} \times \mathbb{Z}_n^*$     [$\sigma_A$ is the sampling information of $A$]

**Remark 4.4.** From the value params, the pair $(h, \beta) \in \mathbb{G}$ such that its sampling information $\alpha \in \mathbb{Z}_n^*$ is unknown (see section 4.1).

A high level description of the Compute algorithm is given below.

A-3.
> **Compute**
> *Input:* (master-key, params, $A, B$), where $A, B \in G_1 \times \mathbb{Z}_{n^2}^*$
> *Step-1.* Use master-key $= \lambda$ to decide if $(A, B) \stackrel{?}{\in} \mathbb{G}^2$ [See section 4.3, item 8]
> *Step-2.* If $(A, B) \notin \mathbb{G}^2$, set $C \leftarrow I \in \mathbb{G}$; otherwise, compute $A \star B$ using $\lambda$ and set $C \leftarrow A \star B$ [See section 4.3, item 7]
> *Output:* $C \in \mathbb{G}$

**Functionality Of Oracle $\mathcal{O}$:** Access to Compute is provided in a black-box manner via the oracle $\mathcal{O}$ that knows master-key and params. The oracle works as follows.

> **Oracle $\mathcal{O}$**
>
> *Input:* $A, B \in G_1 \times \mathbb{Z}_{n^2}^*$
>
> *Step-1.* Set $C \leftarrow \mathsf{Compute}(\text{master-key}, \text{params}, A, B)$
>
> *Output:* $C \in \mathbb{G}$ [We say $C = \mathcal{O}(A, B)$]

**Remark 4.5.** A query to oracle $\mathcal{O}$ on inputs $(A, B) \notin \mathbb{G}^2$ requires at most two exponentiations in $G_1$ and $\mathbb{Z}_{n^2}^*$. On the other hand, if $(A, B) \in \mathbb{G}^2$, the query always involves three exponentiations in $G_1$ and $\mathbb{Z}_{n^2}^*$. Also, $\mathcal{O}(A, B) = A \star B$ whenever $(A, B) \in \mathbb{G}^2$.

**Remark 4.6.** Assuming that access to oracle $\mathcal{O}$ is authentic, we can use $\mathcal{O}$ to decide if any given pair $(x, y) \stackrel{?}{\in} \mathbb{G}$. Additionally we can use $\mathcal{O}$ to compute $A^i$ for any $A \in \mathbb{G}, i \in \mathbb{N}$ using the "repeated squaring and multiply" method [25, p.71].

Since access to oracle $\mathcal{O}$ is over an insecure public channel, we cannot assume that oracle replies are authentic. Denote by $\mathcal{O}^*$ the unauthenticated oracle (which could be an active adversary) supposedly claiming to be oracle $\mathcal{O}$.

The following algorithm Verify-In-Group uses oracle $\mathcal{O}^*$ to decide that any given pair $(x, y) \in G_1 \times \mathbb{Z}_{n^2}^*$ is indeed an element of $\mathbb{G}$. If $(x, y) \notin \mathbb{G}$ the algorithm outputs 0 with a high probability.

A-4.
> **Verify-In-Group**
>
> *Input:* (params, $x, y$) such that $(x, y) \in G_1 \times \mathbb{Z}_{n^2}^*$
>
> *Step-1.* Generate $u_1, u_2, v_1, v_2, \stackrel{R}{\leftarrow} \mathbb{Z}_n$ and $w_1, w_2 \stackrel{R}{\leftarrow} \mathbb{Z}_n^*$
>
> *Step-2.* Set $x_1 \leftarrow x^{u_1} g^{v_1} \in G_1$; $x_2 \leftarrow x^{u_2} g^{v_2} \in G_1$
>
> *Step-3.* Set $y_1 \leftarrow y^{u_1} t^{v_1} w_1^n \bmod n^2$; $y_2 \leftarrow y^{u_2} t^{v_2} w_2^n \bmod n^2$; result $\leftarrow 0$
>
> *Step-4.* Set $(x', y') \leftarrow \mathcal{O}^*((x_1, y_1), (x_2, y_2))$
>
> *Step-5.* If $\hat{e}(x', g) = \hat{e}(x_1, x_2)$, set result $\leftarrow 1$
>
> *Output:* result $\in \{0, 1\}$

We prove in appendix A that the above algorithm is sound (under a non-standard assumption). That is, if $(x, y) \notin \mathbb{G}$ then the algorithm outputs 0 with a high probability. However, the converse is not true. Hence, the above algorithm cannot be used to conclude that $(x, y) \notin \mathbb{G}$ if the output is 0.

In some cases, we may need to decide with certainty that a given pair $(x, y)$ is indeed not an element of $\mathbb{G}$. The next algorithm Verify-Not-In-Group enables us to do this using oracle $\mathcal{O}^*$. If $(x, y) \in \mathbb{G}$ the algorithm outputs 0 with a high probability.

A-5.
> **Verify-Not-In-Group**
>
> *Input:* (params, $x, y$) such that $(x, y) \in G_1 \times \mathbb{Z}_{n^2}^*$
>
> *Step-1.* Set a security parameter $j$ and generate a $j$-bit string $a \stackrel{R}{\leftarrow} \{0, 1\}^j$. Set result $\leftarrow 0$. Initialize another $j$-bit string $b \in \{0, 1\}^j$.
>
> *Step-2.* Repeat for $i$ from 1 to $j$ (denote by $a_i$ and $b_i$, the $i^{th}$ bits of $a$ and $b$ respectively).
>
>     i. If $a_i = 1$, set $(x', y') \stackrel{R}{\leftarrow} \mathsf{Sample}(\text{params})$; otherwise, set $(x', y') \leftarrow (x, y)$
>
>     ii. Set $b_i \leftarrow \mathsf{Verify\text{-}In\text{-}Group}(\text{params}, x', y')$
>
> *Step-3.* If $(a = b)$, set result $\leftarrow 1$
>
> *Output:* result $\in \{0, 1\}$

The following lemma shows that the Verify-Not-In-Group algorithm is sound if the Verify-In-Group algorithm is sound.

**Lemma 4.1.** *If the Verify-In-Group algorithm is sound then the Verify-Not-In-Group algorithm is also sound.*

*Proof.* We must show that if the Verify-Not-In-Group algorithm outputs 1 then $(x, y) \notin \mathbb{G}$.

If $(x, y) \in \mathbb{G}$, then $(x', y')$ in step 2 of Verify-Not-In-Group is always an element of $\mathbb{G}$. Now assume that the Verify-In-Group algorithm is sound. Thus, the probability that $a_i = b_i$ is $\frac{1}{2}$ for any $i$. Also, each bit $a_i$ is independent of other bits. Thus, for a total of $j$ bits, $\Pr[(a_i = b_i) \forall 1 \leq i \leq j] = \frac{1}{2^j}$. In other words, if $(x, y) \in \mathbb{G}$ the probability that the Verify-Not-In-Group algorithm outputs 1 is $\frac{1}{2^j}$, which can be made arbitrarily small. □

The next algorithm Verify takes as input a 3-tuple $(A, B, C)$, where $A, B \in \mathbb{G}$ and $C \in G_1 \times \mathbb{Z}_{n^2}^*$. It outputs 1 only if $C = A \star B$

A-6.
> **Verify**
> 
> *Input:* (params, $A, B, C$) such that $A, B \in \mathbb{G}$ and $C \in G_1 \times \mathbb{Z}_{n^2}^*$.
> Assume that the input is correct.
> 
> *Step-1.* Set $(x_A, y_A) \leftarrow A$; $(x_B, y_B) \leftarrow B$; $(x_C, y_C) \leftarrow C$; result $\leftarrow 0$
> 
> *Step-2.* If $\hat{e}(x_C, g) = \hat{e}(x_A, x_B)$, set result $\leftarrow$ Verify-In-Group(params, $x_C, y_C$)
> 
> *Output:* result $\in \{0, 1\}$

Clearly, the Verify algorithm is sound if the Verify-In-Group algorithm is sound. We observe that we can remove the function call Verify-In-Group(params, $x_C, y_C$) in step 2 of the above algorithm (and simply set result $\leftarrow 1$ instead) without introducing any weakness in the construction. However, including this call enables us to reduce the soundness of other related algorithms to the soundness of the Verify-In-Group algorithm.

Algorithm V-Compute takes as input two elements $A, B \in \mathbb{G}$. It uses the Verify-In-Group algorithm as a subroutine and computes $A \star B$ verifiably by querying $\mathcal{O}^*$.

A-7.
> **V-Compute**
> 
> *Input:* (params, $A, B$) such that $A, B \in \mathbb{G}$. Assume that the input is correct.
> 
> *Step-1.* Set $C \leftarrow \mathcal{O}^*(A, B) \in G_1 \times \mathbb{Z}_{n^2}^*$
> 
> *Step-2.* If Verify$(A, B, C) = 0$, set $C \leftarrow I \in \mathbb{G}$
> 
> *Output:* $C \in \mathbb{G}$

Clearly, the soundness of the above algorithm reduces to the soundness of the Verify algorithm. As a consequence, we state the following theorem which says that if the Verify algorithm is sound then having indirect access to the oracle $\mathcal{O}$ via some active adversary $\mathcal{O}^*$ is the same has having authentic and public access to $\mathcal{O}$.

**Theorem 4.2.** *If the Verify algorithm is sound then $\mathcal{O}$ is a V-Oracle.*

The next algorithm, TD-Exponentiate ("trapdoor-exponentiate") takes as input (i) the sampling information $\sigma_A \in \mathbb{Z}_n^*$ of an element $A \in \mathbb{G}$, (ii) an arbitrary index $i \in \mathbb{Z}$, and (iii) an element $B \in \mathbb{G}$. It outputs $A^i \star B \in \mathbb{G}$. TD-Exponentiate will be primarily used as a subroutine in the Blind and Unblind algorithms.

> **TD-Exponentiate**
>
> *Input:* (params, $\sigma_A, i, B$), where $\sigma_A \in \mathbb{Z}_n^*$; $i \in \mathbb{Z}$; $B \in \mathbb{G}$.
> Here, $\sigma_A$ is the sampling information of $A \in \mathbb{G}$. Assume that the input is correct.
>
> A-8.
>
> *Step-1.* Generate $r \xleftarrow{R} \mathbb{Z}_n^*$
>
> *Step-2.* Set $\sigma \leftarrow \sigma_A{}^i \in \mathbb{Z}_n^*$; $(x_B, y_B) \leftarrow B \in G_1 \times \mathbb{Z}_{n^2}^*$
>
> *Step-3.* Set $x \leftarrow x_B{}^\sigma \in G_1$; $y \leftarrow (y_B)^\sigma r^n = \mathbf{E}(\sigma \mathbf{D}(y_B) \bmod n) \in \mathbb{Z}_{n^2}^*$
>
> *Output:* $(x, y) \in \mathbb{G}$

The next two algorithms Blind and Unblind work as follows.

Blind takes as input a value $A \in \mathbb{G}$. It generates $B \xleftarrow{R} \mathbb{G}$ and outputs $(A \star B) \in \mathbb{G}$, along with $\sigma_B \in \mathbb{Z}_n^*$, the sampling information of $B$. Unblind is the inverse of Blind. It takes as input an pair $(A, \sigma_B) \in \mathbb{G} \times \mathbb{Z}_n^*$ and outputs $A \star B^{-1} \in \mathbb{G}$ such that $\sigma_B$ is the sampling information of $B \in \mathbb{G}$.

> **Blind**
>
> *Input:* (params, $A$) such that $A \in \mathbb{G}$. Assume that the input is correct.
>
> A-9.
>
> *Step-1.* Set $(B, \sigma_B) \xleftarrow{R}$ Sample(params) $\in \mathbb{G} \times \mathbb{Z}_n^*$ [$B$ will be ignored]
>
> *Step-2.* Set $(x, y) \leftarrow$ TD-Exponentiate(params, $\sigma_B, 1, A$) $\in \mathbb{G}$
>
> *Output:* $(x, y, \sigma_B) \in \mathbb{G} \times \mathbb{Z}_n^*$

> **Unblind**
>
> *Input:* (params, $A, \sigma_B$), where $A \in \mathbb{G}$ and $\sigma_B \in \mathbb{Z}_n^*$.
> Here, $\sigma_B$ is the sampling information of $B \in \mathbb{G}$. Assume that the input is correct.
>
> A-10.
>
> *Step-1.* Set $(x, y) \leftarrow$ TD-Exponentiate(params, $\sigma_B, -1, A$) $\in \mathbb{G}$
>
> *Output:* $(x, y) \in \mathbb{G}$

**Lemma 4.3.** *The* Blind/Unblind *algorithms provide information theoretic secrecy.*

*Proof.* Clearly, the Blind and Unblind algorithms are inverses of each other. Now, if the output of the Sample algorithm is uniformly distributed over $\mathbb{G}$ then the output of the Blind algorithm is also uniformly distributed over $\mathbb{G}$, independent of the input. □

Algorithm PV-Compute takes as inputs $A, B \in \mathbb{G}$. It uses the Blind, Unblind and and V-Compute algorithms as subroutines to compute $A \star B$ privately and verifiably.

> **PV-Compute**
>
> *Input:* (params, $A, B$) such that $A, B \in \mathbb{G}$. Assume that the input is correct.
>
> *Step-1.* Set $(A', \sigma_{A'}) \xleftarrow{R}$ Blind(params, $A$) $\in \mathbb{G} \times \mathbb{Z}_n^*$
>
> A-11.
>
> *Step-2.* Set $(B', \sigma_{B'}) \xleftarrow{R}$ Blind(params, $B$) $\in \mathbb{G} \times \mathbb{Z}_n^*$
>
> *Step-3.* Set $C' \leftarrow$ V-Compute$(A', B') \in G_1 \times \mathbb{Z}_{n^2}^*$
>
> *Step-4.* Set $C \leftarrow$ Unblind(params, Unblind(params, $C', \sigma_{A'}), \sigma_{B'}) \in \mathbb{G}$
>
> *Output:* $C \in \mathbb{G}$

Since the Blind/Unblind algorithms provide information theoretic secrecy (lemma 4.3), the soundness of the above algorithm also reduces to the soundness of the Verify algorithm. As a consequence, we state the following theorem which says that if the Verify algorithm is sound then having indirect access to the oracle $\mathcal{O}$ via some active adversary $\mathcal{O}^*$ is the same has having private and authentic access to $\mathcal{O}$.

**Theorem 4.4.** *If the Verify algorithm is sound then $\mathcal{O}$ is a PV-Oracle.*

This completes our O-SAOWF construction. Figure 5 gives the dependencies between the eleven algorithms. We can essentially use PV-Compute$(A, B)$ to denote $f(A, B)$, where $f$ is a real SAOWF defined using $(\mathbb{G}, \star)$ in section 2.1. When considering the security, we will assume that $\mathcal{O}$ takes one time unit to respond to each query and that the sum of the number of queries to $\mathcal{O}$ and the running time of an adversary attacking the O-SAOWF is bounded by a polynomial in $\tau$.

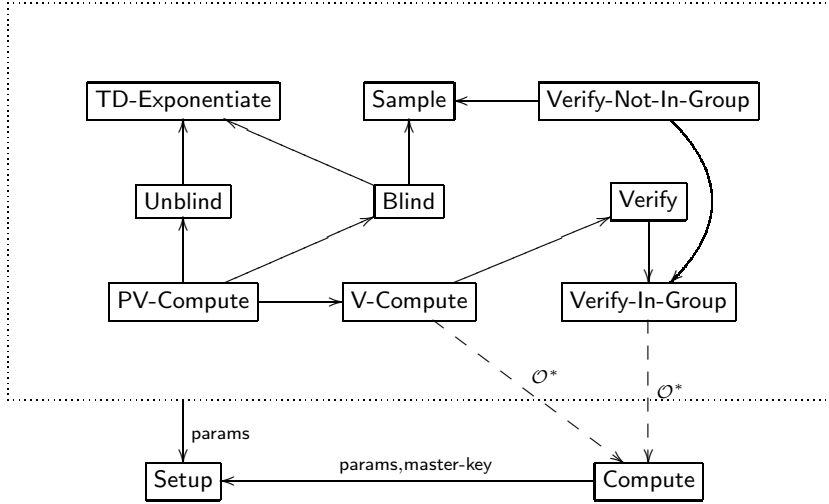

Figure 5: Dependencies between the algorithms.

## 4.5 Notation

For convenience we will adopt the following shorthand notation.

1. We will denote TD-Exponentiate(params, $\sigma_A, i, B$) by $\mathcal{T}(\sigma_A, i, B)$.

2. Since invoking V-Compute is equivalent to making a public query to oracle $\mathcal{O}$ (Theorem 4.2), we will denote V-Compute(params, $A, B$) simply by $\mathcal{O}(A, B)$.

3. Invoking PV-Compute is equivalent to making a private query to oracle $\mathcal{O}$ (Theorem 4.4). We will denote PV-Compute(params, $A, B$) by $\widehat{\mathcal{O}}(A, B)$.

4. For any set of $k$ elements $\{A_1, A_2, \ldots A_k\} \subset \mathbb{G}$, we denote by $\left\langle \mathcal{O} \right\rangle_{i=1}^{k} (A_i)$ the value

$$\mathcal{O}(\mathcal{O}(\ldots \mathcal{O}(A_1, A_2), \ldots), A_k) = \prod_{i=1}^{k} A_i$$

Similarly, we denote by $\left\langle \widehat{\mathcal{O}} \right\rangle_{i=1}^{k} (A_i)$ the value

$$\widehat{\mathcal{O}}(\widehat{\mathcal{O}}(\ldots \widehat{\mathcal{O}}(A_1, A_2), \ldots), A_k) = \prod_{i=1}^{k} A_i$$

5. We will denote by $\mathcal{E}(A, i)$ an algorithm to compute $A^i$ for any $A \in \mathbb{G}$ with the repeated squaring method using V-Compute as a subroutine. This algorithm does not provide privacy of inputs. However, the outputs are verifiable.

6. We will denote by $\widehat{\mathcal{E}}(A, i)$ an algorithm to compute $A^i$ for any $A \in \mathbb{G}$ with the repeated squaring method using PV-Compute as a subroutine. This algorithm provides information theoretic privacy of inputs and verifiability of outputs.

**Remark 4.7.** Computing $A^i$ using algorithms $\mathcal{E}$ and $\widehat{\mathcal{E}}$ will amount to $\approx c \cdot log(i)$ queries to oracle $\mathcal{O}$ (for constant $c$) with the repeated squaring method [25, p.71].

## 4.6 Security Of The Construction

The oracle is primarily used as a "computing device" in the proofs. We assume that the oracle always functions correctly and keeps the trapdoor information $\lambda$ secret. Recall that out of params, the pair $(h, \beta) \in \mathbb{G}$. Denote this value by $P$. The security of our O-SAOWF relies on the difficulty of inverting $\star$ with respect to $P$. One way to do this would be to extract $\lambda$ from the oracle. However, this is equivalent to factoring $n$ so we should look at indirect methods for inverting $\star$ (with respect to $P$) using the oracle. The security of all our constructions reduces to the difficulty of the following problem:

**Group Inversion Problem [GIP$_\mathbb{G}$]:** Let $P = (h, \beta) \stackrel{R}{\leftarrow} \mathbb{G}$ be uniformly sampled using secret $\alpha \stackrel{R}{\leftarrow} \mathbb{Z}_n^*$ such that $h = g^\alpha \in G_1$ and $\beta = \mathbf{E}(\alpha) \in \mathbb{Z}_{n^2}^*$. Given $P$, compute $P^{-1} = (h', \beta') \in \mathbb{G}$, where $h' = g^{1/\alpha} \in G_1$ and $\beta' = \mathbf{E}(1/\alpha) \in \mathbb{Z}_{n^2}^*$, possibly by using the oracle $\mathcal{O}$.

Computing $h'$ becomes an instance of the inverse Diffie-Hellman problem IDHP$_{(g,G_1)}$ defined in section 3.1, which is believed to be hard even if the Diffie-Hellman problem is easy. We hypothesize that any method of reducing IDHP$_{(g,G_1)}$ to CDHP$_{(g,G_1)}$ will yield a method of reducing GIP$_\mathbb{G}$ to the oracle $\mathcal{O}$. We define the advantage of an algorithm for solving the group inversion problem as follows.

**Definition 4.8.** *For any algorithm $\mathcal{A}$, the advantage of $\mathcal{A}$ in solving the group inversion problem GIP-$Adv_\mathcal{A}(\tau)$ for some security parameter $\tau$ is defined as:*

$$GIP\text{-}Adv_\mathcal{A}(\tau) = \Pr\left[\begin{array}{c} \mathcal{A}^{\mathcal{O}(\lambda)}(\hat{e}, G_1, G_2, n, g, t, h, \beta) = (g^{1/\alpha}, \mathbf{E}(1/\alpha)) : \\ \\ (\hat{e}, G_1, G_2, p, q) \stackrel{R}{\leftarrow} \mathcal{BDH}(\tau) \text{ s.t. } |G_1| = |G_2| = pq, \\ n = pq, \quad \alpha \stackrel{R}{\leftarrow} \mathbb{Z}_n^*, \quad g \stackrel{R}{\leftarrow} G_1 \text{ s.t. } \langle g \rangle = G_1, \\ t \stackrel{R}{\leftarrow} \mathbb{Z}_{n^2}^* \text{ s.t. } |\langle t \rangle| = n\lambda, \quad h = g^\alpha, \quad \beta = \mathbf{E}(\alpha) \end{array}\right] \quad (8)$$

*Here $\mathcal{BDH}$ is the BDH parameter generator algorithm (section 3.1.2); $\mathbf{E}$ denotes the Paillier encryption algorithm with public key $(t, n)$ (section 3.2), and $\mathcal{O}$ is an oracle implementing the Compute algorithm (section 4.4).*

For any algorithm $\mathcal{A}$, let $\delta_\mathcal{A}$ denote the upper-bound on the running time of $\mathcal{A}$, and let $k_{(\mathcal{O},\mathcal{A})}$ denote the upper-bound on the number of queries to oracle $\mathcal{O}$ by $\mathcal{A}$. Our security is based on the following conjuncture.

**Conjuncture 4.9.** *For any algorithm $\mathcal{A}$ such that $k_{(\mathcal{O},\mathcal{A})}, \delta_\mathcal{A} \in Poly(\tau)$, GIP-$Adv_\mathcal{A}(\tau)$ is a negligible function in $\tau$. In other words, for all $k_\mathcal{O}, \delta, 1/\epsilon \in Poly(\tau)$, the O-SAOWF is $(k_\mathcal{O}, \delta, \epsilon)$-secure under an adaptive attack using definition 2.5.*

# 5 Applications Of O-SAOWFs

In this section we describe three applications of O-SAOWFs: (a) Multiparty-Key Agreement, (b) Signatures and (c) Broadcast encryption (another application, Identity Based Encryption (IBE) is described in appendix B).

## 5.1 Key Generation (Setup PKI)

To participate in the protocols, each user $i$ must have a certified public key and the corresponding private key. This is generated as follows. Recall that out of params, the pair $(h, \beta) = P \in \mathbb{G}$. This will serve as a common starting value for all users.

1. User $i$ generates $(X_i, \sigma_{X_i}) \xleftarrow{R} \mathsf{Sample}(\mathsf{params}) \in \mathbb{G} \times \mathbb{Z}_n^*$. The private key is $\sigma_{X_i}$.

2. User $i$ computes the public key $Y_i \leftarrow \mathcal{T}(\sigma_{X_i}, 1, P) = X_i \star P$. The public key is made available in an authentic way.

## 5.2 Multiparty Key Agreement

In this section, we describe the multiparty key agreement protocol of Rivest, Rabi and Sherman [6] using O-SAOWFs. At a high level, the objective of a multiparty key agreement protocol is to enable a set of users to compute a shared secret key (the *group private key*) such that no one outside the set can compute this key. In our model each group private key also has a corresponding *group public key*, which can be used for join/merge operations and for verifying (group) signatures created using the group private key. Our construction also defines a *partial public key* that is used in the intermediate steps for group private key computation.

### 5.2.1 Key Agreement Protocol

[$k$ **users**] A set $s = \{1, 2, 3 \ldots k\}$ of $k$ users compute a shared group key.

1. *Partial public key:* Each user $j \in s$ first computes the partial public key

$$Y_{s \setminus \{j\}} \leftarrow \left\langle \mathcal{O} \right\rangle_{i=1; i \neq j}^{k} (Y_i) = \prod_{i=1; i \neq j}^{k} Y_i = P^{k-1} \star \prod_{i=1; i \neq j}^{k} X_i$$

2. *Group Private Key:* Each user $j \in s$ then computes the group private key

$$K_s \leftarrow \mathcal{T}(\sigma_{X_j}, 1, Y_{s \setminus \{j\}}) = X_j \star Y_{s \setminus \{j\}} = P^{k-1} \star \prod_{i=1}^{k} X_i$$

3. *Group Public Key:* The group public key for $s$ is computed by anyone as

$$Y_s \leftarrow \left\langle \mathcal{O} \right\rangle_{i=1}^{k} (Y_i) = \prod_{i=1}^{k} Y_i = P^k \star \prod_{i=1}^{k} X_i$$

Thus, the partial public key of user $j$ in set $s$ is the group public key of the set $s \setminus \{j\}$.

### 5.2.2 Overview Of The Key Agreement Protocol

1. *Complexity*: For a group of $k$ users, $k - 2$ oracle queried are required for each user to compute the shared key. Thus, total $k(k-2)$ queries are required for all the $k$ users. However, no specific ordering is required between the users (users can compute the shared key *after* receiving a ciphertext). Additionally, oracle queries can be batched.

2. *Universal Escrow*: Given a public key $Y_i = X_i \star P$, the oracle $\mathcal{O}$ can compute the corresponding private key $\sigma_{X_i}$. Therefore, $\mathcal{O}$ has universal escrow capability.

3. *Non-interactivity*: Assuming that all the public keys $Y_i$ are known in advance, any user can compute the shared key without interacting with the other users.

4. *Multiple copies of the Oracle*: An arbitrary number of "copies" of the oracle can be run without any compromise in security.

### 5.2.3 Join And Merge Operations

Clearly, members can join any group and many groups can merge arbitrarily. For simplicity we only demonstrate the merge operation between two disjoint sets $a$ and $b$ of users.

**Example [Merge]** A set $a$ of users merges with another set $b$ of users such that $a \cap b = \emptyset$. Further assume that $a$ has the private key $K_a$ and the public key $Y_a$. Similarly, $b$ has the private key $K_b$ and the public key $Y_b$

1. *Group private key:* Each member $i \in a$ computes $K_{a \cup b} \leftarrow \widehat{\mathcal{O}}(K_a, Y_b)$, while each member $j \in b$ computes $K_{a \cup b} \leftarrow \widehat{\mathcal{O}}(K_b, Y_a)$.

2. *Group public key:* The group public key corresponding to the group private key $K_{a \cup b}$ can be computed as $Y_{a \cup b} \leftarrow \mathcal{O}(Y_a, Y_b) = Y_a \star Y_b$.

In the above merge procedure, we assumed that $a$ and $b$ are disjoint (i.e. they have no common members). In case the sets are not disjoint, we could still use the above merge procedure without any serious drawback as long as this instantiation of O-SAOWF is only used for key agreement (and not for signatures, which are discussed below in section 5.3). In case the same instantiation of O-SAOWF is also used for signatures, we would require the merge procedure to eliminate duplicate users in the merged set (this can be efficiently done if the intermediate values in the partial public key computation are cached).

### 5.2.4 Forward Secrecy

Due to the above mentioned merge procedure, the compromise of the group private key of a set $a$ of users compromises the group private key of any other set $c$ of users whenever $c \supsetneq a$. To overcome this weakness, if the private key of group $a$ is compromised, at least one member of $a$ must compute a new public-private key pair. Compromise of a group private key of a set $a$ of users, however, does not compromise the group private key of any set $c$ of users whenever $c \subsetneq a$.

### 5.2.5 Security Of The Key Agreement Protocol

From the key agreement procedure, it is clear that if the adversary knows the private key of user $i \in a$ then the adversary knows the group private key of the set $a$ of users. Additionally, if the adversary knows the group private key of the set $a$ then the adversary also knows the group private key of any set that properly includes $a$. Thus, we restrict the adversary to output the private key of any set $a$ of users such that the adversary knows neither the group private key of any proper subset of $a$ nor the private keys of any users in the set $a$. We show that any algorithm that breaks the above key agreement protocol (with the above restriction) can be used to compute $P^{-1}$. First observe that the secret key $K_a$ for the set $a = \{1, 2, \ldots k\}$ of users is related to the public keys $\{Y_1, Y_2, \ldots Y_k\}$ as:

$$K_a = P^{k-1} \prod_{i=1}^{k} X_i = P^{-1} \star \prod_{i=1}^{k} Y_i \tag{9}$$

We use the security model of security of multiparty key agreement similar to the one used in [7], namely security under a *one-time key attack*. The difference here is that we allow the attacker to choose the set of public keys to attack. Formally, we define a one-time key attack on a multiparty key agreement using game 1.

*Game 1*

**Initialize.** To initialize the game, the challenger gives a security parameter $\tau$ to the adversary. The adversary $\mathcal{A}$ responds with a value $\mu_1 \in \mathbb{N}$

**Challenge.** The challenger $\mathcal{C}$ performs the key generation phase and gives a set $\{Y_1, Y_2, \ldots Y_{\mu_1}\}$ of $\mu_1$ public keys to $\mathcal{A}$.

**Output.** Eventually $\mathcal{A}$ outputs a pair $\langle a, K_a \rangle$.

*Result:* $\mathcal{A}$ wins the game if $a \subseteq \{1, 2, \ldots \mu_1\}$ and $K_a$ is the group private key of $a$.

**Definition 5.1.** *We say that adversary $\mathcal{A}$ $(\mu_1, \delta_1, \epsilon_1)$-breaks the key agreement protocol in an* one-time key attack *if for a total of $\mu_1$ public keys output in the setup phase $\mathcal{A}$ runs at most time $\delta_1$ and the probability of $\mathcal{A}$ winning game 1 is at least $\epsilon_1$. Alternatively we say that the key agreement protocol is $(\mu_1, \delta_1, \epsilon_1)$-secure under a one-time key attack if no such adversary exists.*

The next theorem shows that the key agreement protocol is secure under a one-time key attack if the group inversion problem is hard.

**Theorem 5.1.** *Let the O-SAOWF be $(\cdot, \delta, \epsilon)$-secure under an adaptive attack. Then the multiparty key agreement protocol is $(\mu_1, \delta_1, \epsilon_1)$-secure in a one-time key attack, where $\delta \leq \delta_1 + \Theta(c_1 \mu_1)$; and $\epsilon = \epsilon_1$. Here, $c_1$ is the time for a multiplication in $\mathbb{Z}_n^*$.*

*Proof.* Let the O-SAOWF be $(\cdot, \delta, \epsilon)$-secure under an adaptive attack and let $\mathcal{A}$ be an algorithm that $(\mu_1, \delta_1, \epsilon_1)$-breaks the key agreement protocol in a one-time key attack. We construct an algorithm $\mathcal{B}$ that uses $\mathcal{A}$ to solve $\text{GIP}_{\mathbb{G}}$ in at most $\delta$ time with probability at least $\epsilon$, thus arriving at a contradiction. The input to $\mathcal{B}$ is $P \in \mathbb{G}$ and its goal is to output $P^{-1}$. $\mathcal{B}$ simulates the challenger of game 1 and runs algorithm $\mathcal{A}$ as follows.

**Initialize.** $\mathcal{B}$ gives the security parameter $\tau$ to $\mathcal{A}$ who replies with $\mu_1$.

**Challenge.** $\mathcal{B}$ generates $(Y_1, \sigma_{Y_1}), (Y_2, \sigma_{Y_2}), \ldots (Y_{\mu_1}, \sigma_{Y_{\mu_1}}) \xleftarrow{R} \mathsf{Sample}(\mathsf{params}) \in \mathbb{G} \times \mathbb{Z}_n^*$ and gives the $(\mu_1 + 1)$-tuple $(Y_1, Y_2, \ldots Y_{\mu_1}, P)$ to $\mathcal{A}$.

**Output.** Eventually $\mathcal{A}$ outputs a pair $\langle a, K_a \rangle$.

*Result:* If $\langle a, K_a \rangle$ is a winning configuration, then $a \subseteq \{1, 2, \ldots \mu_1\}$ and $K_a = P^{-1} \star \prod_{i \in a} Y_i$ by virtue of equation 9. Algorithm $\mathcal{B}$ then proceeds as follows:

(a) If $\langle a, K_a \rangle$ is not a winning configuration, $\mathcal{B}$ reports failure and terminates.

(b) We know that $\langle a, K_a \rangle$ is a winning configuration. Algorithm $\mathcal{B}$ sets $\sigma_Y \leftarrow \prod_{i \in a} \sigma_{Y_i} \bmod n$. Thus, $\sigma_Y$ is the sampling information of $\prod_{i \in a} Y_i$ (see section 4.3, item 3).

(c) $\mathcal{B}$ sets $\mathsf{result} \leftarrow \mathcal{T}(\sigma_Y, -1, K_a)$ and outputs $\mathsf{result}$.

Algorithm $\mathcal{B}$ is correct because

$$\mathcal{T}(\sigma_Y, -1, K_a) = \left(\prod_{i \in a} Y_i\right)^{-1} \star K_a = \left(\prod_{i \in a} Y_i\right)^{-1} \star P^{-1} \star \prod_{i \in a} Y_i = P^{-1}$$

The running time of $\mathcal{B}$ is the running time of $\mathcal{A}$ plus the time required for generating the $\mu_1$ public keys; the time required for computing $\mathcal{T}$; and the time required for at most $\mu_1$ multiplications in $\mathbb{Z}_n^*$. The probability of $\mathcal{B}$'s success is the same as the probability of $\mathcal{A}$'s success. This gives the bounds. □

## 5.3 Signatures

As noted in [10], SAOWFs give rise to signature schemes. Here, we describe two signature schemes using O-SAOWFs: ordinary signatures and multi-user signatures. A signature scheme consists of three algorithms KeyGen, Sign and VerifySig, where the algorithms have their usual constraints [22]. Our message space is $\mathbb{N}$.

### 5.3.1 Single-User Signatures

This is a variation of the scheme for single-user signatures described in [6].

**KeyGen.** This algorithm is described in section 5.1. The private key of user $i$ is $\sigma_{X_i} \in \mathbb{Z}_n^*$. The public key is $Y_i = X_i \star P \in \mathbb{G}$.

**Sign.** Let $m \in \mathbb{N}$ be the message. To sign $m$, user $i$ computes the signature $S_{(i,m)}$ as:

$$S_{(i,m)} \leftarrow \mathcal{T}(\sigma_{X_i}, m, P) = X_i^m \star P$$

**VerifySig.** To verify a signature $S_{(i,m)}$ of user $i$ on message $m$, we check if the following holds:

$$\mathcal{E}(Y_i, m) \stackrel{?}{=} \mathcal{O}(S_{(i,m)}, \mathcal{E}(P, m-1))$$

In other words, we check if $Y_i^m \stackrel{?}{=} S_{(i.m)} \star P^{m-1}$

### 5.3.2 Multi-User And Ring Signatures

The above construction of single-user signatures can be trivially extended to multi-user signatures. To sign messages, members of a group must share a secret group key.

**KeyGen.** This algorithm is described in section 5.2. Without loss of generality, assume that any of the set $a = \{1, 2, \ldots j\}$ of users want to independently sign messages using the group private key $K_a = P^{j-1} \star \prod_{i=1}^{j} X_i$ such that the signatures can be verified using the group public key $Y_a = \prod_{i=1}^{j} Y_i$.

**Sign.** Let $m \in \mathbb{N}$ be the message. To sign $m$, any member $i \in a$ computes the signature $S_{(a,m)}$ as:

$$S_{(a,m)} \leftarrow \widehat{\mathcal{O}}(\widehat{\mathcal{E}}(K_a, m), P) = K_a^m \star P$$

**VerifySig.** To verify a signature $S_{(a,m)}$ of user $i \in a$ on message $m$, we check if the following holds:

$$\mathcal{E}(Y_a, m) \stackrel{?}{=} \mathcal{O}(S_{(a,m)}, \mathcal{E}(P, m-1))$$

In other words, we check if $Y_a^m \stackrel{?}{=} S_{(a,m)} \star P^{m-1}$

Given a signature of some set $a$, it is not possible for any group controller to revoke the anonymity of the signer (since there is no group controller). Thus, the above scheme is an example of ring signatures [34].

### 5.3.3 Security Of The Signature Schemes

The strongest model for security of signatures is security against *existential forgery* under an *adaptive chosen message attack* [22], where the attacker is required to output a successful forgery under the challenge public key after having access to the signing oracle. However, we only prove the security of our schemes in a weaker model that we call security against *existential forgery* under a *non-adaptive chosen message attack*. In a non-adaptive attack, the attacker is not allowed to make any signature queries. We define this using the following game between the challenger $\mathcal{C}$ and an adversary $\mathcal{A}$.

*Game 2*

**Initialize.** To initialize the game, the challenger gives a security parameter $\tau$ to the adversary. The adversary $\mathcal{A}$ outputs $\mu_2 \in \mathbb{N}$.

**Challenge.** The challenger $\mathcal{C}$ performs the key generation phase and gives a set $\{Y_1, Y_2, \ldots Y_{\mu_2}\}$ of $\mu_2$ public keys to $\mathcal{A}$.

**Output.** Eventually $\mathcal{A}$ outputs a tuple $\langle a, S_{(a,m)}, m \rangle$.

*Result:* $\mathcal{A}$ wins the game if $a \subseteq \{1, 2, \ldots \mu_2\}$ and $S_{(a,m)}$ is a valid signature by $a$ on the message $m$.

**Definition 5.2.** *We say that adversary $\mathcal{A}$ $(\mu_2, \delta_2, \epsilon_2)$-breaks the signature scheme in a non-adaptive chosen message attack if for a total of $\mu_2$ public keys output in the setup phase $\mathcal{A}$ runs at most time $\delta_2$ and the probability of $\mathcal{A}$ winning game 2 is at least $\epsilon_2$. Alternatively we say that the signature scheme is $(\mu_2, \delta_2, \epsilon_2)$-secure under a non-adaptive chosen message attack if no such adversary exists.*

The next theorem shows that any algorithm that is successful in existential forgery of signatures under a non-adaptive chosen message attack can be used to compute $P^{-1}$. First observe that $S_{(a,m)}$ can be rewritten as

$$S_{(a,m)} = K_a{}^m \star P = P^{1-m} \star (\prod_{i=1}^{j} Y_i)^m \tag{10}$$

Also note that Game 2 considers both single and multi-user signatures.

**Theorem 5.2.** *Let the O-SAOWF be $(k_{\mathcal{O}}, \delta, \epsilon)$-secure under an adaptive attack. Then the signature scheme is $(\mu_2, \delta_2, \epsilon_2)$-secure under a non-adaptive chosen message attack, where $k_{\mathcal{O}} \leq c_2 \log(n)$; $\delta \leq \delta_2 + \Theta(\mu_2)$; and $\epsilon \geq \epsilon_2$. Here, $c_2$ is a constant.*

*Proof.* Let the O-SAOWF be $(k_{\mathcal{O}}, \delta, \epsilon)$-secure under an adaptive attack and let $\mathcal{A}$ be an algorithm that $(\mu_2, \delta_2, \epsilon_2)$-breaks the signature scheme in a non-adaptive chosen message attack. We construct an algorithm $\mathcal{B}$ that uses $\mathcal{A}$ to solve $\text{GIP}_{\mathbb{G}}$ in at most $\delta$ time with probability at least $\epsilon$, thus arriving at a contradiction. The input to $\mathcal{B}$ is $P \in \mathbb{G}$ and its goal is to output $P^{-1}$. $\mathcal{B}$ simulates the challenger of game 2 and runs algorithm $\mathcal{A}$.

 **Initialize.** $\mathcal{B}$ gives the parameter $\tau$ to $\mathcal{A}$, who outputs $\mu_2 \in \mathbb{N}$.

 **Challenge.** $\mathcal{B}$ generates $(Y_1, \sigma_{Y_1}), (Y_2, \sigma_{Y_2}), \ldots (Y_j, \sigma_{Y_{\mu_2}}) \xleftarrow{R} \mathsf{Sample}(\mathsf{params}) \in \mathbb{G} \times \mathbb{Z}_n^*$ and gives the $(\mu_2 + 1)$-tuple $(Y_1, Y_2, \ldots Y_{\mu_2}, P)$ as the input to $\mathcal{A}$.

 **Output.** Finally $\mathcal{A}$ outputs a tuple $\langle a, S_{(a,m)}, m \rangle$.

 *Result:* If the tuple $\langle a, S_{(a,m)}, m \rangle$ represents a winning configuration, then $a \subseteq \{1, 2, \ldots \mu_2\}$ and $S_{(a,m)} = P^{1-m} \star (\prod_{i \in a} Y_i)^m$ by virtue of equation 10. Algorithm $\mathcal{B}$ then proceeds as follows:

 (a) If $\langle a, S_{(a,m)}, m \rangle$ not a winning configuration, algorithm $\mathcal{B}$ reports failure and terminates.

 (b) We know that $a \subseteq \{1, 2, \ldots \mu_2\}$ and $S_{(a,m)} = P^{1-m} \star (\prod_{i \in a} Y_i)^m$. Algorithm $\mathcal{B}$ then sets $C \leftarrow \mathcal{E}(P, m-2) = P^{m-2}$ and $\sigma_Y \leftarrow \prod_{i \in a} \sigma_{Y_i} \bmod n$. Thus, $\sigma_Y$ is the sampling information of $\prod_{i \in a} Y_i$ (see section 4.3, item 3).

 (c) Finally, $\mathcal{B}$ sets $\mathsf{result} \leftarrow \mathcal{T}(\sigma_Y, -m, \mathcal{O}(S_{(a,m)}, C))$ and outputs $\mathsf{result}$.

Algorithm $\mathcal{B}$ is correct because

$$\mathcal{T}(\sigma_Y, -m, \mathcal{O}(S_{(a,m)}, C)) = (\prod_{i \in a} Y_i)^{-m} \star S_{(a,m)} \star C$$

$$= (\prod_{i \in a} Y_i)^{-m} \star (P^{1-m} \star \prod_{i \in a} Y_i) \star (P^{m-2}) = P^{-1}$$

The running time of $\mathcal{B}$ is the running time of $\mathcal{A}$ plus the time required for generating the $\mu_2$ public keys; the time required for computing $\mathcal{T}$; and the time required for at most $\mu_2$ multiplications in $\mathbb{Z}_n^*$. The probability of $\mathcal{B}$'s success is the same as the probability of $\mathcal{A}$'s success. Finally, $\mathcal{B}$ queries the oracle for computing $\mathcal{O}(S_{(a,m)}, C)$ and $\mathcal{E}(P, m-2)$. This amounts to a maximum of $c_2 \log(n)$ queries for some constant $c_2$. Thus, we have the required bounds. □

## 5.4 Broadcast Encryption

In a broadcast encryption scheme [35], anyone can encrypt a message addressed to a closed set of users using their public keys such that only those users have the ability to decrypt the message (we do not consider schemes that allow *traitor tracing* [36]). Using our method, the size of ciphertexts and public/private keys is $O(1)$ and for a set of $k$ users, a total of $O(k)$ calls to the oracle $\mathcal{O}$ are required for encryption and decryption. A broadcast encryption scheme consists of three algorithms KeyGen, BC-Encrypt and BC-Decrypt, where the algorithms have their usual constraints [35]. (We use the prefix 'BC' to indicate 'broadcast').

**KeyGen.** This algorithm is described in section 5.2. Without loss of generality, assume that messages will be encrypted to any arbitrary set $a = \{1, 2, \ldots k\}$ of $k$ users with public keys $\{Y_1, Y_2, \ldots Y_k\}$. The sender of the message generates the group public key $Y_a$ by making $k-1$ oracle queries as follows:

$$Y_a \leftarrow \langle \mathcal{O} \rangle_{i=1}^{k} (Y_i) = \prod_{i=1}^{k} Y_i = P^k \star \prod_{i=1}^{k} X_i$$

and any receiver $i \in a$ must independently compute the group private key $K_a$ by making $k-2$ oracle queries as follows:

$$K_a \leftarrow \mathcal{T}(\sigma_{X_j}, 1, \prod_{i=1; i \neq j}^{k} Y_i) = P^{k-1} \star \prod_{i=1}^{k} X_i$$

We additionally require a cryptographic hash function $\mathcal{H}_1 : G_1 \mapsto \{0,1\}^l$, that will be treated as a random oracle in the proofs.[7] Our message space is $\{0,1\}^l$.

**BC-Encrypt.** To encrypt $m \in \{0,1\}^l$ to the set $a = \{1, 2, \ldots k\}$ of $k$ users with group public key $Y_a$, generate $(R, \sigma_R) \xleftarrow{R} \mathsf{Sample}(\mathsf{params}) \in \mathbb{G} \times \mathbb{Z}_n^*$ and compute

$$c_1 \leftarrow m \oplus \mathcal{H}(\mathcal{T}(\sigma_R, 1, Y_a)) = m \oplus \mathcal{H}(R \star Y_a)$$

$$C_2 \leftarrow \mathcal{T}(\sigma_R, 1, P) = R \star P$$

Here $\oplus$ denotes the XOR operator. The ciphertext is $C = (c_1, C_2) \in \{0,1\}^l \times \mathbb{G}$.

**BC-Decrypt.** To decrypt ciphertext $(c_1, C_2)$ using group private key $K_a$, compute

$$m \leftarrow c_1 \oplus \mathcal{H}(\widehat{\mathcal{O}}(K_a, C_2)) = c_1 \oplus \mathcal{H}(K_a \star C_2)$$

The decryption is correct, because for a legitimate ciphertext we have

$$(K_a \star C_2) = (P^{k-1} \star \prod_{i=1}^{k} X_i) \star (R \star P) = R \star P^k \star \prod_{i=1}^{k} X_i = R \star Y_a$$

### 5.4.1 Security Of Broadcast Encryption

We use a restricted model for security called security under an adaptive chosen plaintext attacks (IND-CPA). In this model, we fix some arbitrary set $a = \{1, 2, \ldots k\}$ of $k$ users and require the adversary to attack the semantic security of the scheme without access to a decryption oracle. However, we allow the adversary to choose the subset of keys it is attacking. Since full security in the sense of adaptive chosen ciphertext attacks (IND-CCA) in the random oracle model can be achieved using the Fujisaki-Okamoto transformation [37], we only prove security in the IND-CPA model. IND-CPA security of a broadcast encryption scheme is defined using the following game between a challenger $\mathcal{C}$ and an adversary $\mathcal{A}$.

---

[7]To construct this hash function, let $A = (x, y) \in \mathbb{G} \in G_1 \times \mathbb{Z}_n^*$ be some input and let $\mathcal{H}_1 : G_1 \mapsto \{0,1\}^l$ be a hash function. Then $\mathcal{H}(A) = \mathcal{H}_1(x)$.

*Game 3*

**Initialize.** The challenger $\mathcal{C}$ gives a security parameter $\tau$ to the adversary $\mathcal{A}$, who outputs a tuple $\mu_3$. The challenger performs the key generation phase and gives a set $\{Y_1, Y_2, \ldots Y_{\mu_3}\}$ of $\mu_3$ public keys to $\mathcal{A}$.

**Challenge.** $\mathcal{A}$ generates two messages $m_0, m_1$ along with a set $a \subseteq \{1, 2, \ldots \mu_3\}$ of users. The challenger chooses a bit $b \stackrel{R}{\leftarrow} \{0, 1\}$ and outputs the encryption of $m_b$ under the group public key $Y_a$ of $a$.

**Guess.** Eventually $\mathcal{A}$ outputs a bit $b' \in \{0, 1\}$

*Result:* $\mathcal{A}$ wins the game if $b = b'$.

We refer to such an adversary $\mathcal{A}$ as an IND-CPA adversary. We define $\mathcal{A}$'s advantage in attacking the broadcast encryption scheme Adv-cpa$_\mathcal{A}(\tau)$ as:

$$\text{Adv-cpa}_\mathcal{A}(\tau) = \left| \Pr[b = b'] - \frac{1}{2} \right|,$$

where the probability is taken over the random coin tosses of $\mathcal{C}$ and $\mathcal{A}$.

**Definition 5.3.** *Let $\mathcal{H}$ be a random oracle. We say that an IND-CPA adversary $\mathcal{A}$ $(\mu_3, \delta_3, k_3, \epsilon_3)$-breaks the broadcast encryption scheme in a adaptive chosen plaintext attack if for a total of $\mu_3$ public keys output in the setup phase $\mathcal{A}$ runs at most time $\delta_3$; $\mathcal{A}$ makes at most $k_3$ queries to the oracle for $\mathcal{H}$; and Adv-cpa$_\mathcal{A}(\tau)$ at least $\epsilon_3$. Alternatively we say that the broadcast encryption scheme is $(\mu_3, \delta_3, k_3, \epsilon_3)$-secure under a adaptive chosen plaintext attack if no such adversary $\mathcal{A}$ exists.*

The next theorem shows that any IND-CPA adversary $\mathcal{A}$ with non-negligible advantage Adv-cpa$_\mathcal{A}(\tau)$ in the random oracle model can be used to solve the group inversion problem with non-negligible advantage. The proof is similar to the proof of [23, lemma 4.3]

**Theorem 5.3.** *Let $\mathcal{H}$ be a random oracle and let the O-SAOWF be $(\cdot, \delta, \epsilon)$-secure under an adaptive attack. Then the broadcast encryption scheme is $(\mu_3, \delta_3, k_3, \epsilon_3)$-secure under an adaptive chosen plaintext attack, where $\delta \leq \delta_3 + \Theta(c_1 \mu_3) + \Theta(c_2 k_3)$; and $\epsilon \geq 2 \cdot \epsilon_3$. Here, $c_1$ is the time for one multiplication in $\mathbb{Z}_n^*$, and $c_2$ is a constant that depends on the oracle $\mathcal{O}$.*

*Proof.* Let the O-SAOWF be $(\cdot, \delta, \epsilon)$-secure under an adaptive attack and let $\mathcal{A}$ be an algorithm that $(\mu_3, \delta_3, k_3, \epsilon_3)$-breaks the key agreement protocol in an adaptive chosen plaintext attack. We construct an algorithm $\mathcal{B}$ that uses $\mathcal{A}$ to solve GIP$_\mathbb{G}$ in at most $\delta$ time with probability at least $\epsilon$, thus arriving at a contradiction. The input to $\mathcal{B}$ is $P \in \mathbb{G}$ and its goal is to output $P^{-1}$. $\mathcal{B}$ simulates the challenger of game 3 and runs $\mathcal{A}$.

**Initialize.** $\mathcal{B}$ gives the security parameter $\tau$ to $\mathcal{A}$ who replies with $\mu_3$. $\mathcal{B}$ generates

$$(Y_1, \sigma_{Y_1}), (Y_2, \sigma_{Y_2}), \ldots (Y_{\mu_3}, \sigma_{Y_{\mu_3}}) \stackrel{R}{\leftarrow} \textsf{Sample}(\textsf{params}) \in \mathbb{G} \times \mathbb{Z}_n^*,$$

and gives the $(\mu_3 + 1)$-tuple $(Y_1, Y_2, \ldots Y_{\mu_3}, P)$ to $\mathcal{A}$.

**$\mathcal{H}$-queries.** At any time, $\mathcal{A}$ may query the random oracle $\mathcal{H}$. To respond to these queries, $\mathcal{B}$ maintains a list of tuples called the $\mathcal{H}^{list}$. Each entry in this list is a tuple of the form $\langle Z_j, \mathcal{H}_j \rangle$. Initially this list is empty. To respond to a $\mathcal{H}$ query on $Z_i$, algorithm $\mathcal{B}$ does the following:

  (a) If the query $Z_i$ already appears on the $\mathcal{H}^{list}$ in a tuple $\langle Z_i, \mathcal{H}_i \rangle$, then $\mathcal{B}$ responds with $\mathcal{H}(Z_i) = \mathcal{H}_i$.
  (b) Otherwise, $\mathcal{B}$ just picks a random string $\mathcal{H}_i \in \{0, 1\}^l$ and adds the tuple $\langle Z_i, \mathcal{H}_i \rangle$ to the $\mathcal{H}^{list}$. It responds with $\mathcal{H}(Z_i) = \mathcal{H}_i$.

**Challenge.** $\mathcal{A}$ generates two messages $m_0, m_1$ along with a set $a \subseteq \{1, 2, \ldots \mu_3\}$ and sends the tuple $(m_0, m_1, a)$ to $\mathcal{B}$. Algorithm $\mathcal{B}$ picks random $c_1 \in \{0,1\}^l$; generates $(C_2, \sigma_{C_2}) \xleftarrow{R} \mathsf{Sample}$; defines the $C = (c_1, C_2)$; and gives $C$ as the challenge ciphertext to $\mathcal{A}$. Observe that the decryption of $C$ is $c_1 \oplus \mathcal{H}(P^{-1} \star C_2 \star \prod_{i \in a} Y_i)$.

Algorithm $\mathcal{B}$ also computes $\sigma_W \leftarrow \sigma_{C_2} \cdot \prod_{i \in a} \sigma_{Y_i} \bmod n$. Clearly, $\sigma_W$ is the sampling information of $W = C_2 \star \prod_{i \in a} Y_i$ (see section 4.3, item 3).

**Guess.** Eventually $\mathcal{A}$ outputs a bit $b' \in \{0, 1\}$. At this point, $\mathcal{B}$ searches the $\mathcal{H}^{list}$ to find a tuple $\langle Z_j, \mathcal{H}_j \rangle$ such that

$$\mathcal{O}(Z_j, P) = W \tag{11}$$

If such a tuple does not exist in the $\mathcal{H}^{list}$, algorithm $\mathcal{B}$ reports failure and terminates. Otherwise, $\mathcal{B}$ sets $\mathsf{result} \leftarrow \mathcal{T}(\sigma_W, -1, Z_j) = W^{-1} \star Z_j$. Algorithm $\mathcal{B}$ outputs $\mathsf{result}$ as the solution to the $\mathrm{GIP}_\mathbb{G}$ instance.

Clearly, the simulation provided by algorithm $\mathcal{B}$ is sound. Therefore, from claims 1 and 2 in the proof of [23, lemma 4.2], we can conclude that

$$\Pr\left[\text{a tuple } \langle Z_j, \mathcal{H}_j \rangle \text{ appears in the } \mathcal{H}^{list} \text{ such that equation 11 is satisfied}\right] \geq 2 \cdot \epsilon_3$$

Thus, $\epsilon \geq 2 \cdot \epsilon_3$. The running time of $\mathcal{B}$ is the running time of $\mathcal{A}$ plus the time required for generating the $\mu_3$ public keys; the time required for computing $\mathcal{T}$; the time required for searching up to $k_3$ entries in the $\mathcal{H}^{list}$; and the time required for at most $\mu_3$ multiplications in $\mathbb{Z}_n^*$. Thus, $\delta \leq \delta_3 + \Theta(c_1 \mu_3) + \Theta(c_2 k_3)$, where $c_1$ is the time for one multiplication in $\mathbb{Z}_n^*$, and $c_2$ is the time for checking one entry of the $\mathcal{H}^{list}$. Thus, we have the required bounds □

# 6 Implementation And Efficiency

In this section, we will briefly touch upon issues relating to implementation and efficiency of our primitive. Although our construction of O-SAOWF has other applications as demonstrated, we feel that its primary use will be for highly dynamic group key agreement in applications like "secure chat". Our system offers the advantage that the group key need not be precomputed for communication between group members. Thus, there is no specific ordering between the users.

## 6.1 Key Size

Factoring $n$ enables an attacker to solve $\mathrm{GIP}_\mathbb{G}$. Based on the current state of the art factoring algorithms, we suggest using the modulus $n$ of about 313 decimal digits ($\approx 1024$ bits) for moderate security applications.[8] This also makes computing discrete logarithms in $G_1$ intractable using Pollard's rho method [25, p.128]. Using these parameters elements of $\mathbb{G}$ can be represented with at most $\approx 384$ bytes. The public keys $Y_i$ of section 5.1, which are elements of $\mathbb{G}$ will be 384 bytes each. The private keys $\sigma_{X_i}$ on the other hand, which are elements of $\mathbb{Z}_n^*$ will be 128 bytes.

## 6.2 Query Overhead

In all the above protocols, we have been working in the equivalence classes of $\mathbb{G}$ rather than the individual elements themselves. For any $A = (x, y) \in \mathbb{G}$, the equivalence class $[A]$ is completely characterized by the first element $x$. The second element $y$ is used only as an 'auxiliary' input for the oracle, and is useless to anyone who does not know the factorization of $n$. Thus, verification of the second element cannot provide additional security. With this consideration in mind, we slightly modify the Verify algorithm of section 4.4 and remove the call to the Verify-In-Group subroutine, since computing the bilinear pairing allows verification of the first element $x$. The computation overhead is given in table 1.

---

[8]See the RSA factoring challenge (http://www.rsasecurity.com/rsalabs/node.asp?id=2092) and the article "TWIRL and RSA key size" (http://www.rsasecurity.com/rsalabs/node.asp?id=2004). It is thought that 1024 bit keys will be secure till the year 2010 while 2048 bit keys will be secure till the year 2030.

| Algorithm | Exp $G_1$ | Exp $\mathbb{Z}_{n^2}^*$ | Multi $\mathbb{Z}_{n^2}^*$ | Multi $\mathbb{Z}_n^*$ | Pairing |
|---|---|---|---|---|---|
| Compute | 3 | 4 | 1 | 2 | - |
| V-Compute | 3 | 4 | 1 | 2 | 1 |
| PV-Compute | 5 | 5 | 1 | 2 | 1 |

Table 1: Computation involved in a query

## 6.3 Batch Queries

For increased efficiency in partial public key computation, we will assume that calls to the oracle can be batched as follows, for any $i$ inputs $A_1, A_2, \ldots A_i \in \mathbb{G}$, the oracle outputs $A_1 \star A_2 \star \ldots A_i$. In this case, for key computation in a group of $m$ users each user must make a batch call requiring a message of size $O(m)$ bits to be sent to the oracle. The reply of the oracle constitutes just one element of size $O(1)$. However, we lose the ability to verify the output of the oracle in a "batch query".

## 6.4 Verifiability Of The Oracle

If verifiability of the oracle is not required (i.e. we need protection only from passive adversaries) then instead of the bilinear group $G_1$, we can use a finite field having a multiplicative subgroup of order $n$. The set $\mathbb{S}$ defined in section 4.2 is then the $\phi(n)$ elements of this field of order $n$.

## 6.5 Fast Paillier Decryption

Since each computation of $\star$ requires two decryptions, it is desirable to obtain a faster decryption procedure. In [29, section 6], a fast variant of the Paillier cryptosystem is presented where decryption does not require the factors of $n$ and runs with almost quadratic complexity. In this variant, $\lambda = (p-1)(q-1)$ has a large prime factor $\nu$. The public key is $(t, n)$ such that the order of $t \in \mathbb{Z}_{n^2}^*$ is $\nu n$. The private key is $\nu$. Encryption and decryption is described below.

**Encrypt** Plaintext is $m \in \mathbb{Z}_n$. Generate $r \xleftarrow{R} \mathbb{Z}_n^*$ and compute $c = t^{m+nr} \mod n^2$. The ciphertext is $c$.

**Decrypt** Ciphertext is $c \in \mathbb{Z}_{n^2}^*$. Compute $m = \frac{L(c^\nu \mod n^2)}{L(t^\nu \mod n^2)} \mod n$.

Semantic security of this variant does not depend on the DCRA assumption (section 3.2) but instead relies on the weaker *Decisional Partial Discrete Logarithm Assumption* (DPDLA) [29, theorem 20], which states that the following problem is hard.

**Decisional Partial Discrete Logarithm Problem (DPDLP$_{(t,n)}$)** Fix any $t \in \mathbb{Z}_{n^2}^*$ such that the order of $t$ is $\nu n$ for unknown $\nu$. Given $w \in \langle t \rangle$ and $x \in \mathbb{Z}_n$, output 1 if $\exists y \in \mathbb{Z}_n^*$ s.t $w \equiv t^x y^n \pmod{n^2}$ otherwise output 0.

## 6.6 Decentralizing The Oracle

Distributing the oracle is desirable, since each oracle call involves 3 exponentiations in $G_1$ (irrespective of the decryption algorithm). It is possible to share the Paillier decryption key (known only to the oracle) between different trusted authorities with the weakness that compromise of even one would compromise the entire system. We close this section with a comparison of our scheme with previously proposed group key agreement methods in table 2.

# 7 Conclusion

In this paper, we presented a practical implementation of a new cryptographic primitive known as an Oracle Strong Associative One-Way Function (O-SAOWF). As some practical applications of this primitive, we presented a one-round key agreement scheme for dynamic ad-hoc groups based on the protocol due to Rabi and Sherman [6].

| Membership size is $m$ | **O-SAOWF** | **GDH** basic [5] | **AGKE** [38] | **GKE**[39] |
|---|---|---|---|---|
| Number of rounds | 1 | $m-1$ sequential | 2 sequential | 2 sequential |
| Synchronization ordering needed? | No | Yes | Yes | Yes |
| Controller needed? | No | No | Yes (initial key distribution) | Yes (group key distribution) |
| Interaction needed? | No | Yes | Yes | Yes (for synchronization) |
| Key Agreement method | Oracle | Self (interactive) | Self (broadcast) | Controller |
| Message size per user (sent)* | $(m-1)k_1$ | $(m-1)k_2$ | $k_3$ (broadcast only, otherwise $mk_3$ | $2k_4$ (to controller) |
| Message size per user (rcvd)* | $k_1$ (no verification), otherwise $(m-2)k_1$ | $(m-1)k_2$ | $mk_3$ | $k_4$ |
| Merge with $m_1$ users | 1 round (total $2(m+m_1)$ messages) | $O(m+m_1)$ rounds (fresh key) | 2 rounds (total $m+m_1$ broadcasts) | 2 rounds (total $2m_1+m$ messages) |
| Part with $m_1$ users | no oracle calls needed if partial keys are cached | $O(m-m_1)$ rounds (fresh key) | 2 rounds ($m-m_1$ to $m+m_1$ broadcasts) | 1 round (total $m-m_1$ messages) |
| Partial Public keys reusable ? | Yes | No | No | No |
| Optimization Possible? | Yes** | Not likely | Not likely | Not likely |
| Protection under active attack | Yes# (Verifiable Oracle) | Susceptible to a man-in-the-middle attack | Authentication after 2nd round | Insecure under an active attack [40] |
| Protection under passive attack | Group Inversion Problem | Diffie-Hellman Problem | Diffie-Hellman Problem | Diffie-Hellman Problem |

  * We assume that $k_1, k_2 \ldots k_n$ are constants.

  ** Assuming that intermediate controllers are used and partial public keys are cached.

  # If public keys are known in advance, the verifiability of the oracle ensures *implicit group key authentication*.

Table 2: Comparison of our group key agreement scheme

The scheme can be extended to group signatures as demonstrated in section 5.3. In reality, we also demonstrate a "pay-per-use" cryptographic primitive using the oracle. The advantage of our scheme in comparison with other centralized schemes is that the central controller does not maintain any state information of the groups it is managing. It just acts as a "computing device" for users registered with it. We envisage several interesting applications of this primitive in the near future.

As we demonstrate, the ability to "multiply" using the oracle does not give us the ability to "divide" in $\mathbb{G}$ because its order is unknown. This ensures that an "Euclidean"-like Algorithm does not work here. The curious property of our O-SAOWF is that it is *weakly invertible*. In other words, given $A \in \mathbb{G}$, it is possible to compute two pairs $(B, B') \in \mathbb{G}^2$ such that $A = B \star B'$ even without using the oracle.[9] We conclude this section with two open questions.

1. Prove/disprove conjuncture 4.9. In other words, find the complexity of the group inversion problem $\text{GIP}_{\mathbb{G}}$.

---

[9]To see this, sample $B \leftarrow \mathbb{G}$. Then $B' = B^{-1} \star A$.

2. Construct a group of hidden order where the group operation is computable and strongly non-invertible. In other words, exhibit a practical SAOWF construction.

# Acknowledgment

We would like to thank Ronald Rivest, Virendra Sule, Pascal Paillier and Chunbo Ma for useful feedback during the preparation of this manuscript.

# References

[1] Whitfield Diffie and Martin E. Hellman. New directions in cryptography. *IEEE Transactions on Information Theory*, IT-22(6):644–654, 1976.

[2] Antoine Joux. A one round protocol for tripartite Diffie-Hellman. In *ANTS-IV: Proceedings of the 4th International Symposium on Algorithmic Number Theory*, pages 385–394, London, UK, 2000. Springer-Verlag.

[3] Sandro Rafaeli and David Hutchison. A survey of key management for secure group communication. *ACM Comput. Surv.*, 35(3):309–329, 2003.

[4] Xukai Zou, Byrav Ramamurthy, and Spyros S. Magliveras. *Secure Group Communications Over Data Networks*. Springer, New York, NY, USA, 2005.

[5] Michael Steiner, Gene Tsudik, and Michael Waidner. CLIQUES: A new approach to group key agreement. In *Proceedings of the 18th International Conference on Distributed Computing Systems (ICDCS'98)*, pages 380–387, Amsterdam, 1998. IEEE Computer Society Press.

[6] Muhammad Rabi and Alan T. Sherman. An observation on associative one-way functions in complexity theory. *Inf. Process. Lett.*, 64(5):239–244, 1997.

[7] Dan Boneh and Alice Silverberg. Applications of multilinear forms to cryptography. Cryptology ePrint Archive, Report 2002/080, 2002.

[8] Alan T. Sherman. *Cryptology and VLSI (a two-part dissertation). I, II, Detecting and exploiting algebraic weaknesses in cryptosystems. Algorithms for placing modules on a custom VLSI chip*. Thesis (Ph.D.), Laboratory for Computer Science, Massachusetts Institute of Technology, Cambridge, MA, USA, October 1986. Supervised by Ronald Linn Rivest.

[9] Burton S. Kaliksi, Jr., Ronald L. Rivest, and Alan T. Sherman. Is the Data Encryption Standard a group? *Journal of Cryptology*, 1(1):3–36, 1988.

[10] M. Rabi and A. Sherman. Associative one-way functions: A new paradigm for secret-key agreement and digital signatures. Technical Report CS-TR-3183/UMIACS-TR-93-124, 1993.

[11] Christopher M. Homan. Low ambiguity in strong, total, associative, one-way functions, 2000.

[12] Alina Beygelzimer, Lance A. Hemaspaandra, Christopher M. Homan, and Jörg Rothe. One-way functions in worst-case cryptography: Algebraic and security properties. Technical Report TR722, 1999.

[13] Lane A. Hemaspaandra and Jörg Rothe. Creating strong, total, commutative, associative one-way functions from any one-way function in complexity theory. *J. Comput. Syst. Sci.*, 58(3):648–659, 1999.

[14] Lane A. Hemaspaandra, Jörg Rothe, and Amitabh Saxena. Enforcing and defying associativity, commutativity, totality, and strong noninvertibility for one-way functions in complexity theory. In *ICTCS*, 2005.

[15] Susan Hohenberger. The cryptographic impact of groups with infeasible inversion. Master's thesis, Massachusetts Institute of Technology, 2003. Advisor: Ronald L. Rivest.

[16] Amitabh Saxena and Ben Soh. A novel method for authenticating mobile agents with one-way signature chaining. In *Proceedings of The 7th International Symposium on Autonomous Decentralized Systems (ISADS 05)*, pages 187–193, China, 2005. IEEE Computer Press.

[17] Lane A. Hemaspaandra, Kari Pasanen, and Jörg Rothe. If $P \neq NP$ then some strongly noninvertible functions are invertible. In *FCT '01: Proceedings of the 13th International Symposium on Fundamentals of Computation Theory*, pages 162–171. Springer-Verlag, 2001.

[18] Ueli Maurer. Information-theoretic cryptography. In Michael Wiener, editor, *Advances in Cryptology — CRYPTO '99*, volume 1666 of *Lecture Notes in Computer Science*, pages 47–64. Springer-Verlag, August 1999.

[19] David Chaum. Blind signatures for untraceable payments. In *CRYPTO'82*, pages 199–203, 1982.

[20] László Babai and Endre Szemerédi. On the complexity of matrix group problems I. In *FOCS'1984*, pages 229–240, 1984.

[21] Ronald L. Rivest. On the notion of pseudo-free groups. In Moni Naor, editor, *TCC*, volume 2951 of *Lecture Notes in Computer Science*, pages 505–521. Springer, 2004.

[22] Dan Boneh, Ben Lynn, and Hovav Shacham. Short signatures from the Weil pairing. In *ASIACRYPT '01: Proceedings of the 7th International Conference on the Theory and Application of Cryptology and Information Security*, pages 514–532, London, UK, 2001. Springer-Verlag.

[23] Dan Boneh and Matthew K. Franklin. Identity-based encryption from the Weil pairing. *SIAM J. Comput.*, 32(3):586–615, 2003.

[24] Paulo S. L. M. Barreto, Hae Yong Kim, Ben Lynn, and Michael Scott. Efficient algorithms for pairing-based cryptosystems. In *CRYPTO '02: Proceedings of the 22nd Annual International Cryptology Conference on Advances in Cryptology*, pages 354–368, London, UK, 2002. Springer-Verlag.

[25] Alfred J. Menezes, Scott A. Vanstone, and Paul C. Van Oorschot. *Handbook of Applied Cryptography*. CRC Press, Inc., Boca Raton, FL, USA, 1996.

[26] Antoine Joux and Kim Nguyen. Separating Decision Diffie-Hellman from Diffie-Hellman in cryptographic groups. Technical Report 2001/003, 2001.

[27] Neal Koblitz. *A course in number theory and cryptography.* Springer-Verlag New York, Inc., New York, NY, USA, 1987.

[28] Dan Boneh, Eu-Jin Goh, and Kobbi Nissim. Evaluating 2-DNF formulas on ciphertexts. In Joe Kilian, editor, *TCC*, volume 3378 of *Lecture Notes in Computer Science*, pages 325–341. Springer, 2005.

[29] Pascal Paillier. Public-key cryptosystems based on composite degree residuosity classes. In *EUROCRYPT*, pages 223–238, 1999.

[30] R. L. Rivest, A. Shamir, and L. Adleman. A method for obtaining digital signatures and public-key cryptosystems. *Commun. ACM*, 21(2):120–126, 1978.

[31] M. O. Rabin. Digitalized signatures and public-key functions as intractable as factorization. Technical report, Massachusetts Institute of Technology, Cambridge, MA, USA, 1979.

[32] Dario Catalano, Rosario Gennaro, and Nick Howgrave-Graham. The bit security of Paillier's encryption scheme and its applications. In *EUROCRYPT '01: Proceedings of the International Conference on the Theory and Application of Cryptographic Techniques*, pages 229–243, London, UK, 2001. Springer-Verlag.

[33] D. Catalano, R. Gennaro, and N. H. Graham. Paillier's trapdoor function hides up to O(n) bits. *Journal of Cryptology*, 15(4):251–269, 2002.

[34] Ronald L. Rivest, Adi Shamir, and Yael Tauman. How to leak a secret. *Lecture Notes in Computer Science*, 2248:552–??, 2001.

[35] Amos Fiat and Moni Naor. Broadcast encryption. *Lecture Notes in Computer Science*, 773:480–??, 1994.

[36] Hervé Chabanne, Duong Hieu Phan, and David Pointcheval. Public traceability in traitor tracing schemes. In Ronald Cramer, editor, *EUROCRYPT*, volume 3494 of *Lecture Notes in Computer Science*, pages 542–558. Springer, 2005.

[37] Eiichiro Fujisaki and Tatsuaki Okamoto. Secure integration of asymmetric and symmetric encryption schemes. *Lecture Notes in Computer Science*, 1666:537–554, 1999.

[38] Hyun-Jeong Kim, Su-Mi Lee, and Dong Hoon Lee. Constant-round authenticated group key exchange for dynamic groups. In Pil Joong Lee, editor, *ASIACRYPT*, volume 3329 of *Lecture Notes in Computer Science*, pages 245–259. Springer, 2004.

[39] E. Bresson, O. Chevassut, A. Essiari, and D. Pointcheval. Mutual authentication and group key agreement for low-power mobile devices, 2003.

[40] Junghyun Nam, Seungjoo Kim, and Dongho Won. Attacks on Bresson-Chevassut-Essiari-Pointcheval's group key agreement scheme for low-power mobile devices. Cryptology ePrint Archive, Report 2004/251, 2004.

# APPENDIX

## A Soundness Of Verify-In-Group Algorithm

The reader is referred to section 4.4, algorithm A-4 for the notation used here. First we define the following problem.

**Decision Exponent Class Problem [$\text{DECP}_{(t,n,g,G_1)}$]:** Given $\{t, n, g, G_1\} \subset \text{params}$ and a pair $(x, y) \in G_1 \times \mathbb{Z}_{n^2}^*$, where $x = g^a$ and $y = t^b r^n \mod n^2$ for unknowns $(a, b, r) \in \mathbb{Z}_n \times \mathbb{Z}_n \times \mathbb{Z}_n^*$, output 1 if $[a \equiv b \pmod{p} \veebar a \equiv b \pmod{q}]$, otherwise output 0.

The following theorem shows that the Verify-In-Group algorithm is sound if $\text{DECP}_{(t,n,g,G_1)}$ is intractable.

**Theorem A.1.** *If the decision exponent class problem is hard then the Verify-In-Group algorithm is sound.*

*Proof.* The input to the Verify-In-Group algorithm is $(x, y) \in G_1 \times \mathbb{Z}_{n^2}^*$. We must show that if the algorithm outputs 1 then $(x, y) \in \mathbb{G}$. Let $x = g^a$ and $y = t^b r^n \mod n^2$ for unknowns $(a, b, r) \in \mathbb{Z}_n \times \mathbb{Z}_n \times \mathbb{Z}_n^*$. The transformation of $(x, y)$ to $(x_1, y_1)$ and $(x_2, y_2)$ in step 2 of the algorithm can be denoted by the mapping

$$\begin{aligned} f_1 : \mathbb{Z}_n \times \mathbb{Z}_n \times \mathbb{Z}_n^* &\mapsto G_1 \times \mathbb{Z}_{n^2}^* \\ (u, v, w) &\mapsto (g^{au+v}, t^{bu+v} r^{un} w^n \mod n^2) \end{aligned}$$

Consider the cases when the algorithm outputs 1.

**Case 1.** $[a \equiv b \pmod{p} \wedge a \equiv b \pmod{q}]$: In this case $a = b$ and so $(x, y) \in \mathbb{G}$. Therefore, $f_1(u, v, w) \in \mathbb{G} \; \forall \; u, v, w \in domain(f_1)$. In this case, the output of Verify-In-Group algorithm is consistent with its requirements.

**Case 2.** $[a \not\equiv b \pmod{p} \wedge a \not\equiv b \pmod{q}]$: It is not hard to prove that the mapping $f_1$ is a bijection in this case. Since both sides of $f_1$ have the same number of elements $n^2 \phi(n)$, it is enough to prove that $f_1$ is invertible with respect to every element in $G_1 \times \mathbb{Z}_{n^2}^*$. Let $(g^{a_1}, t^{b_1} r_1^n \mod n^2) \in G_1 \times \mathbb{Z}_{n^2}^*$ be an element of the right side of $f_1$. If a preimage $(u_1, v_1, w_1)$ of $f_1$ exists for this element, then we must have

$$\left. \begin{aligned} a_1 &\equiv au_1 + v_1 \pmod{n} \\ b_1 &\equiv bu_1 + v_1 \pmod{n} \\ r_1 &\equiv r^{u_1} w_1 \pmod{n} \end{aligned} \right\} \quad (12)$$

Clearly equation 12 has a unique solution in $(u_1, v_1, w_1)$ for all $(a_1, b_1, r_1)$ if and only if $(a - b) \in \mathbb{Z}_n^*$. In other words, if and only if $\gcd(a - b, n) = 1$. Note that $\gcd(a - b, n) = 1$ is another way of saying that $[a \not\equiv b \pmod{p} \wedge a \not\equiv b \pmod{q}]$.

Since $f_1$ is a bijection, the distributions $\{(x_1, y_1)\}$ and $\{(x_2, y_2)\}$ are identical to a random distribution in $G_1 \times \mathbb{Z}_{n^2}^*$. If the oracle $\mathcal{O}^*$ can make the algorithm output 1 then we can use $\mathcal{O}^*$ to solve $\text{CDHP}_{(g,G_1)}$ (see section 3.1) as follows:

1. Input is $g, g^{\sigma_1}, g^{\sigma_2}$ and our goal is to output $g^{\sigma_1 \sigma_2}$.
2. Generate $y_1, y_2 \xleftarrow{R} \mathbb{Z}_{n^2}^*$
3. Set $x_1 \leftarrow g^{\sigma_1}$ and $x_2 \leftarrow g^{\sigma_2}$
4. Give $(x_1, y_1), (x_2, y_2)$ as input to oracle $\mathcal{O}^*$ in step 3 of the algorithm instead of the real values.

Since the forged and real distributions of $\{(x_1, y_1)\}$ and $\{(x_1, y_2)\}$ are identical, the oracle $\mathcal{O}^*$ cannot distinguish between the forged and real inputs. Accordingly it will reply with $(x', y')$ such that the algorithm outputs 1 in step 4. In this case $x'$ is the required solution to the $\text{CDHP}_{(g,G_1)}$ instance.

**Case 3.** $[a \equiv b \pmod{p} \wedge a \not\equiv b \pmod{q}]$: (or $\gcd(a-b, n) > 1$ and $a \neq b$)

The probability of a randomly picked pair $(x, y) \in G_1 \times \mathbb{Z}_{n^2}^*$ such that $\gcd(a-b, n) > 1$ and $a \neq b$ is $\frac{p+q-2}{pq}$ which can be neglected for large $p, q$. On the other hand, if the adversary ($\mathcal{O}^*$) knows in advance that $\gcd(a-b, n) > 1$ but does not know both of $\{a, b\}$, then the adversary knows that the distribution of the image

$$f_1(u_1, v_1, w_1) = (g^{a_1},\ t^{b_1} r_1^n \bmod n^2)$$

always satisfies $a_1 \equiv b_1 \pmod{p}$. In this case, our security relies on the adversary's inability to distinguish elements of this distribution from randomly chosen elements of $G_1 \times \mathbb{Z}_{n^2}^*$ assuming the hardness of $\text{DECP}_{(t,n,g,G_1)}$. Under this assumption, we can use the adversary $\mathcal{O}^*$ to solve $\text{CDHP}_{(g,G_1)}$ as in the previous case. The case of $[a \not\equiv b \pmod{p} \wedge a \equiv b \pmod{q}]$ is handled similarly.

Thus, we have proved that the algorithm is sound under the assumption that the problems $\text{DECP}_{(t,n,g,G_1)}$ and $\text{CDHP}_{(g,G_1)}$ are intractable. □

## B  Identity Based Encryption Using O-SAOWFs

In this section we give (without a security proof) an Identity Based Encryption (IBE) scheme as another application of our O-SAOWFs. We refer the reader to [23] for the definitions of an IBE scheme and to section 4.4 for the notation used here. In summary, out IBE scheme has four PPT algorithms Setup-IBE, KeyGen, ID-Encrypt and ID-Decrypt. The definition of "PPT" has the usual caveat; oracles are considered as algorithms.

1. The Setup-IBE algorithm takes as input some security parameter. It outputs the IBE system parameters par and the IBE master key m-key.

2. The KeyGen algorithm takes as input the value par, m-key and a random string $i$. It outputs the private key prv-key$_i$ corresponding to the string $i$.

3. The ID-Encrypt algorithm takes as input par, a random message $m$ and a random string $i$. It outputs a ciphertext $c$.

4. The ID-Decrypt algorithm takes as input par, a private key prv-key$_i$ (corresponding to some string $i$) and ciphertext $c$. It outputs a message $m$.

The ID-Encrypt and ID-Decrypt algorithms satisfy the standard consistency constraint:

$$\forall m\ \forall i\ \text{ID-Decrypt}(\text{par}, \text{ID-Encrypt}(\text{par}, m, i), \text{KeyGen}(\text{par}, \text{m-key}, i)) = m$$

In an IBE scheme, the master key m-key is known only to a trusted authority known as the Key Generating Center (KGC) that is responsible for distributing private keys. In our construction although the oracle $\mathcal{O}$ is required for computation, it *need not* be the Key Generating Center (KGC). The four algorithms are described below.

1. Setup-IBE: Set $(X, \sigma_X), (Y, \sigma_Y) \xleftarrow{R} \text{Sample}(\text{params})$ and set $Z \leftarrow \mathcal{T}(\sigma_X, 1, Y) = X \star Y$. Finally set par $\leftarrow (Y, Z) \in \mathbb{G}^2$; m-key $\leftarrow (\sigma_X, \sigma_Y) \in \mathbb{Z}_n^{*\,2}$ and output (par, m-key).

2. KeyGen: Let $i \in \mathbb{N}$ be the input string. Set prv-key$_i \leftarrow \mathcal{T}(\sigma_X, -i, Y) = X^{-i} \star Y \in \mathbb{G}$ and output prv-key.

3. ID-Encrypt: Our message space is $\{0,1\}^k$ where $k < \log_2(n)$ and we require a cryptographic hash function $\mathcal{H}: \mathbb{G} \mapsto \{0,1\}^k$. To encrypt a message $m \in \{0,1\}^k$ using input string $i \in \mathbb{N}$, first generate random $(R, \sigma_R) \xleftarrow{R} \text{Sample}(\text{params})$. Then compute

$$c_1 = m \oplus \mathcal{H}(\mathcal{T}(\sigma_R, 1, \mathcal{E}(Y, i+1))) = m \oplus \mathcal{H}(Y^{i+1} \star R)$$

$$C_2 = \mathcal{T}(\sigma_R, 1, \mathcal{E}(Z, i)) = Z^i \star R = X^i \star Y^i \star R$$

The ciphertext is $(c_1, C_2)$.

Both $c_1$ and $C_2$ can be directly computed if $Y^{i+1}$ and $Z^i$ are precomputed.

4. **ID-Decrypt**: To decrypt arbitrary ciphertext $(c_1, C_2)$ compute

$$m = c_1 \oplus \mathcal{H}(\widehat{\mathcal{O}}(C_2, \mathsf{prv\text{-}key}_i)) = c_1 \oplus \mathcal{H}(C_2 \star X^{-i} \star Y)$$

Decryption is correct, because for a legitimate ciphertext:

$$C_2 \star X^{-i} \star Y = (X^i \star Y^i \star R) \star (X^{-i} \star Y) = Y^{i+1} \star R$$